%% file: main_tnsm.tex
\documentclass[journal]{IEEEtran}

\usepackage{booktabs}
\usepackage{multirow}
\usepackage{makecell}
\usepackage{amsmath}
\usepackage{amsfonts}
\usepackage{mathtools}
\usepackage{graphicx}
\usepackage{comment}
\usepackage{xspace} 
\usepackage{xurl}
\usepackage{hyperref}
\usepackage{float}
\usepackage{cite}
\usepackage{amssymb}
\usepackage{subcaption}
\usepackage{listings}
\lstdefinestyle{template_naming}{
  basicstyle=\ttfamily\small,
  breaklines=true,
  frame=single,
  columns=fullflexible
}

\usepackage[table,dvipsnames]{xcolor}

\usepackage[flushleft]{threeparttable}

\usepackage{xcolor}
\usepackage{tabularx}

\definecolor{IndianRed}{RGB}{205,92,92}

\makeatletter
\newcommand{\mysubsubsection}[2]{\@startsection
  {subsubsection}%
  {3}%
  {\parindent}%
  {-1.5ex plus -1ex minus -0.2ex}%
  {0ex plus 0.2ex}%
  {\normalfont\normalsize\itshape}
  {#1}{#2}
  }

\usepackage{enumitem}

\newcommand*{\cloudens}{\text{ClouDens}\@\xspace}
\newcommand*{\LCSs}{\text{LCSs}\@\xspace}
\newcommand*{\LCS}{\text{LCS}\@\xspace}

\newcommand*{\etal}{\textit{et al.}\@\xspace}

\usepackage{tcolorbox}
\tcbuselibrary{skins}

\newcommand{\rqanswer}[2]{%
  \begin{tcolorbox}[
    enhanced,
    colback=white,
    colframe=black!70,
    boxrule=0.6pt,
    arc=4pt,
    left=3pt, right=3pt, top=3pt, bottom=3pt,
  ]
  \textbf{\textit{Answer to #1:}}
  \textit{#2}
  \end{tcolorbox}
}

\newcommand{\rqfirst}{\textit{What is the impact of the context-aware graph on anomaly detection performance in \LCSs?}\@\xspace}

\newcommand{\rqsecond}{\textit{How does anomaly detection performance vary across telemetry subsets and scoring strategies, and what trade-off between anomaly coverage and alert volume does ensembling produce?}\@\xspace}

\newcommand{\rqthird}{\textit{What are the sensitivity and computational costs associated with sliding window size and sparsity imputation?}\@\xspace}

\ifCLASSINFOpdf
\else
\fi

\begin{document}
\title{\cloudens: Operational Context-Aware Anomaly\\ Detection for Large-scale Cloud System Monitoring}

\author{Thu T. H.~Doan, %
        Mohammad~Saiful~Islam, %
        Andriy~Miranskyy, %
        Ngoc-Thanh~Nguyen, \\
        Rogardt~Heldal, %
        and
        Patrizio~Pelliccione %
\thanks{Thu~T.~H.~Doan and Patrizio~Pelliccione are with the Department
of Computer Science, Gran Sasso Science Institute, L'Aquila, Italy 
(e-mail: thihoaithu.doan@gssi.it; patrizio.pelliccione@gssi.it). Patrizio Pelliccione is also Adjunct Professor with the University of Bergen, Norway.} %
\thanks{Mohammad~Saiful~Islam and Andriy~Miranskyy are with the Department
of Computer Science, Toronto Metropolitan University, Toronto, Canada
(e-mail: mohammad.s.islam@torontomu.ca; avm@torontomu.ca).}%
\thanks{Ngoc-Thanh~Nguyen and Rogardt~Heldal are with the Department
of Computer Science, Electrical Engineering and Mathematical Sciences, Western Norway University of Applied Sciences, Bergen, Norway (e-mail: Ngoc.Thanh.Nguyen@hvl.no; Rogardt.Heldal@hvl.no).}%
}

\maketitle

\begin{abstract}
With the rapid growth of cloud computing infrastructures in scale and complexity, network monitoring for Large-scale Cloud Systems (\LCSs) has become increasingly challenging, requiring automated and reliable anomaly detection to maintain service availability. Modern LCSs continuously generate telemetry logs from activities among distributed cloud services, producing high-dimensional multivariate time series that capture system operations. Detecting anomalies in this context remains difficult due to the extreme dimensionality of monitoring telemetry data, complex dependencies among distributed components, and severe sparsity caused by intermittently active services. Taking these challenges into account, we first conduct an empirical study on telemetry logs from the IBM Cloud Console platform for managing \LCSs, and then propose \cloudens, an anomaly detection framework tailored to \LCS monitoring
that leverages operational-context attributes encoded in the telemetry log schema to improve detection accuracy and early identification of anomalies. \cloudens partitions high-dimensional telemetry logs into domain-guided subsets, constructs a context-aware graph to model operational service dependencies, and employs Spatio-Temporal Graph Neural Networks for forecasting-based anomaly detection.

We evaluate \cloudens on the recently released IBM Cloud Telemetry Dataset and provide practical insights into designing reliable anomaly detection solutions for \LCS monitoring. Experimental results demonstrate that \cloudens achieves higher NAB scores in \texttt{count}-based telemetry features, indicating more accurate and earlier anomaly detection with broader anomaly coverage than a GRU-based model. Beyond these improvements, our comprehensive empirical study reveals that telemetry feature subsets, operational-context modeling, scoring strategies, and telemetry sparsity imputation all substantially influence anomaly detection performance. These findings offer practical guidance for designing anomaly detection solutions and for fairly benchmarking existing and future approaches for monitoring \LCSs.

\end{abstract}


\IEEEpeerreviewmaketitle

\section{Introduction \label{sec:introduction}}

\begin{figure}[t]
    \centering
    \includegraphics[width=\columnwidth]{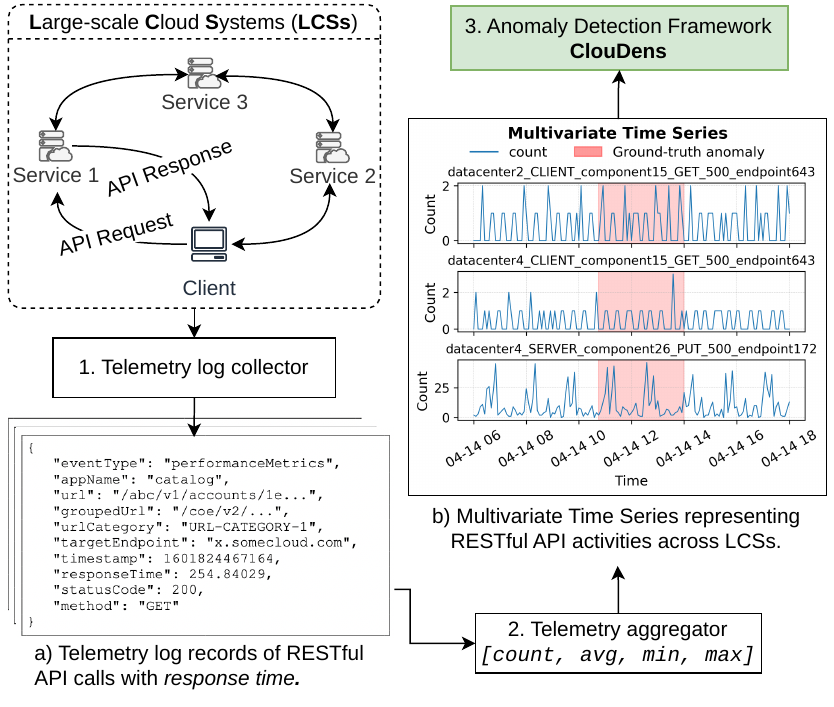}
    \caption{Overview of telemetry-driven anomaly detection for monitoring Large-scale Cloud Systems. Telemetry logs generated by distributed RESTful API calls (example in (a), based on~\cite{islam2021anomaly}) are aggregated into multivariate time series (b), which form the input to forecasting-based detection models.}
    \label{fig:api_logs_to_mtsad}
    \vspace{-15pt}
\end{figure}

Modern cloud platforms have evolved into large-scale distributed systems composed of thousands of interconnected microservices that jointly deliver Internet-scale applications and services~\cite{Xu2017_AdaptiveServiceAPI,HASHEM201598,Hagemann2021_systematic_review,Xin2023ensemble_learning,gartner_public_cloud_2024}. In this paper, we refer to such an environment as a \textit{Large-scale Cloud System} (\LCS/\LCSs), where software and hardware components are deployed across geographically distributed data centers. To ensure service reliability and operational availability, cloud providers continuously collect records of cloud service operations and interactions, comprising RESTful API activities, response time, request counts, and error statistics, among others~\cite{Fielding2000,LERCHER2024112110,Otero2024_distributed_telemetry}. As illustrated in Fig.~\ref{fig:api_logs_to_mtsad}(a), operational activities among distributed services in \LCSs are recorded by a telemetry log collector, which captures RESTful API response time as an example of telemetry logs. These telemetry logs provide a comprehensive view of the operational behavior of cloud services and have become the primary data source for monitoring system health, diagnosing performance degradations, and identifying abnormal operational behaviors, hereafter referred to as \textit{anomalies}.

Although continuous telemetry collection enables real-time anomaly detection~\cite{Toka2021PredictingCloud,islam2021anomaly,islam2025anomaly}, the growing scale and heterogeneity of telemetry logs in \LCSs~\cite{Aceto2013CloudMonitoringSurvey} make manual monitoring and analysis impractical. Consequently, automated anomaly detection has become the main solution for maintaining service reliability, preventing cascading failures, and satisfying Quality-of-Service
(QoS) and Service Level Agreements (SLAs)~\cite{Chkirbene2020,Hrusto2026_monitoring_data_in_cloud_system}. Despite significant advances in machine learning and deep learning techniques, anomaly detection in production environments remains challenging due to the dynamic nature of cloud services, the diversity of telemetry data, and the scale of modern distributed systems~\cite{Miranskyy2016bigdatachallenges,pourmajidi2018challenges,pourmajidi2019dogfoodinguseibmcloud,pourmajidi2021challenging}. These challenges have made anomaly detection for \LCS monitoring an active research area, motivating reliable and scalable methods that explicitly account for the operational characteristics of distributed microservices while remaining practical for real-world deployment.

Recent studies on the IBM Cloud platform have demonstrated the potential of machine learning for automated anomaly detection in \LCSs with cloud telemetry data~\cite{islam2020anomaly,islam2021anomaly}. In particular, Islam \emph{et al.}~\cite{islam2025anomaly} released a publicly available telemetry dataset from the IBM Cloud Console, providing one of the first realistic benchmarks for anomaly detection in \LCSs. The dataset exhibits key characteristics of production cloud telemetry, including high dimensionality, heterogeneous metrics, and severe sparsity. Although baseline methods based on Artificial Neural Networks (ANNs) and Gated Recurrent Units (GRUs) achieve promising results, accurately detecting diverse anomaly patterns without triggering excessive false alarms remains an open challenge, motivating anomaly detection methods that better exploit the structural and operational characteristics of cloud telemetry.

Based on this motivation, we conduct an empirical analysis of the IBM Cloud Telemetry Dataset~\cite{islam2025anomaly} and propose an anomaly detection framework tailored for \LCS monitoring that targets diverse anomaly pattern discovery, early detection, and low false alarms. Our analysis highlights several characteristics of telemetry logs in \LCS monitoring that influence the design of anomaly detection solutions. First, anomalies in production cloud environments are often localized to a subset of services rather than affecting the entire system simultaneously. Second, different groups of telemetry metrics capture complementary aspects of system behavior and therefore exhibit different sensitivities to anomaly patterns. 
These observations suggest that treating all telemetry features uniformly may obscure important operational characteristics and limit anomaly detection performance. Furthermore, when available, operational-context attributes describing cloud service activities can be leveraged to capture realistic operational dependencies, further improving detection accuracy.

In this work, we propose \cloudens, an operational context-aware anomaly detection framework designed for cloud telemetry monitoring in \LCSs. Rather than building a single anomaly detection model from the entire telemetry space, \cloudens partitions telemetry data into domain-guided telemetry subsets that capture different operational aspects of cloud services. For each subset, the framework constructs a context-aware graph that models operational dependencies among monitored service activities and employs Spatio-Temporal Graph Neural Networks (ST-GNNs) for the forecasting-based detection backbone. Finally, the outputs from multiple telemetry subsets are combined through an ensemble strategy to improve anomaly coverage while controlling the increase in alert volume. 
We evaluate \cloudens on the IBM Cloud Telemetry Dataset~\cite{islam2025anomaly}, comparing its performance against a GRU-based model~\cite{cho2014propertiesneuralmachinetranslation} and showing improvements in NAB score, anomaly coverage, and early detection.
Beyond this comparison, we also vary post-processing techniques, including scoring strategies and ensembling, to investigate the trade-off between anomaly coverage and false alarms in \LCS monitoring.

\textbf{Our study does not introduce a new foundation architecture for time series anomaly detection. Instead, we investigate the unique characteristics of telemetry logs and use these insights to design an effective anomaly detection framework for monitoring distributed cloud services.}

Our main contributions are listed below.

\begin{itemize}
\item We propose \cloudens, an operational context-aware anomaly detection framework for \LCS monitoring that integrates domain-guided telemetry feature decomposition, operational-context graph modeling, and spatio-temporal forecasting to improve anomaly detection performance.

\item We conduct an empirical study on the real-world IBM Cloud Telemetry Dataset~\cite{islam2025anomaly} to systematically investigate how telemetry feature subsets, operational-context graph modeling, telemetry sparsity imputation, anomaly scoring strategies, and ensemble configurations influence anomaly detection performance. The resulting findings provide practical guidance for designing reliable anomaly detection frameworks for \LCSs.
\item We release a replication package~\cite{replicationPackage} containing the implementation, experimental results, and evaluation scripts to support reproducibility and future benchmarking of anomaly detection methods for cloud monitoring systems.
\end{itemize}

\noindent\textbf{Paper outline:} 
The paper is organized as follows. Section~\ref{sec:problem_formulation} formulates the problem and describes the IBM Cloud Telemetry Dataset. Section~\ref{sec:related_work} reviews related work on cloud anomaly detection. Section~\ref{sec:cloudens_framework} introduces the \cloudens framework. Section~\ref{sec:evaluation} presents experimental results around three evaluation questions. Section~\ref{sec:threats} discusses practical insights and threats to validity. Section~\ref{sec:conclusion} concludes the paper.

\section{Preliminaries and Problem Formulation\label{sec:problem_formulation}}

This section formulates the anomaly detection problem in the context of \LCS monitoring. Section~\ref{sec:telemetry_logs} introduces the telemetry representation and its characteristics, while Section~\ref{sec:problem_statement} formally defines the problem addressed in this paper.

\subsection{Cloud Telemetry Representation\label{sec:telemetry_logs}}

Cloud providers continuously collect telemetry logs from interactions among distributed microservices to monitor the operational state of \LCSs. Fig.~\ref{fig:api_logs_to_mtsad} illustrates the telemetry monitoring workflow considered in this study. As cloud services communicate through RESTful APIs, each activity generates a telemetry log record containing both operational-context information about the service behaviors and numerical performance measurements. These records are continuously collected and aggregated into multivariate time series using operational metrics such as request count, average response time, minimum response time, and maximum response time, providing the foundation for automated anomaly detection. Throughout this paper, we consider two complementary types of information extracted from telemetry logs associated with RESTful API operations, as detailed below.

\begin{itemize}
    \item[\textbf{(1)}] \textbf{Operational-context attributes}, which describe the semantic and operational characteristics of API activities, including deployment location (e.g., data center or region), service ownership (e.g., microservice or component), and communication role (e.g., \texttt{CLIENT} or \texttt{SERVER}). Additional attributes such as URL patterns, HTTP methods, HTTP status codes, and error types further characterize request semantics and execution outcomes. These metadata describe service dependencies and deployment topology, providing operational-context information. 
    Fig.~\ref{fig:api_logs_to_mtsad}(a) presents an example telemetry log record containing both contextual attributes (e.g., \texttt{appName}, \texttt{targetEndpoint}, \texttt{statusCode}) and numerical performance measurements (e.g., \texttt{responseTime}).

    \item[\textbf{(2)}] \textbf{Numerical performance metrics}, which quantify the operational behavior of cloud services over time and constitute the primary input for anomaly detection. As illustrated in Fig.~\ref{fig:api_logs_to_mtsad}(b), telemetry records are periodically aggregated into operational metrics such as request count, average response time, minimum and maximum response time. These aggregated measurements naturally form Multivariate Time Series (MTS), where each telemetry feature represents the temporal evolution of a specific performance metric associated with an activity of APIs.

\end{itemize}

Together, operational-context attributes and numerical performance metrics provide complementary views of cloud system behavior. Contextual information captures the operational relationships among cloud services, including deployment topology and service dependencies, thereby describing the spatial characteristics. In contrast, numerical performance metrics characterize the temporal evolution of service activity by continuously recording operational measurements over time. Combining these two information sources enables anomaly detection models to jointly exploit spatial dependencies and temporal dynamics for \LCS monitoring.

\textit{\textbf{Example: IBM Cloud Telemetry Dataset. \label{sec:working_dataset}}}
In the context of anomaly detection for \LCS monitoring, we perform a deeper analysis of the publicly available IBM Cloud Telemetry Dataset introduced by Islam \emph{et al.}~\cite{islam2025anomaly}, with its corresponding artifact available on Zenodo~\cite{islam_2024_14062900}. The dataset was collected from the IBM Cloud Console, which serves as the primary management interface for IBM Cloud services. 

\begin{figure*}
\begin{lstlisting}[caption={Pivot dataset column template and column name example in IBM Cloud Telemetry Dataset (see Appendix~B in~\cite{islam2025anomaly}).}, label={lst:feature_template},captionpos=b,style=template_naming,abovecaptionskip=5pt]
Template: {location}_{kind}_{host}_{method}_{statusCode}_{endpoint}_{aggregated_stats_name}
Example: datacenter1_CLIENT_component10_GET_200_endpoint865_count
\end{lstlisting}
\vspace{-20pt}
\end{figure*}

Compared with commonly used public benchmarks such as NAB~\cite{Lavin_2015}, Microsoft~\cite{microsoftCloudDataset}, and Exathlon~\cite{jacob2021exathlon}, the IBM dataset more closely reflects the characteristics of production cloud monitoring. While NAB and Microsoft contain relatively low-dimensional telemetry and Exathlon relies on synthetic anomaly generation, the IBM dataset contains telemetry collected from a real production cloud together with anomaly labels derived from multiple operational sources, including incident reports, testing logs, and operator messages~\cite{islam2025anomaly}. 
Specifically, each telemetry feature is identified by a structured naming convention that encodes contextual information about the monitored API activity, including deployment location, communication role (\texttt{CLIENT}/\texttt{SERVER}), host, HTTP method, HTTP status code, endpoint identifier, and aggregation type (e.g., \texttt{count}, \texttt{avg}, \texttt{min}, and \texttt{max}), as illustrated in Listing~\ref{lst:feature_template}. This representation preserves deployment and service relationships that are later exploited to construct context-aware graphs. Additional details regarding the telemetry schema are available in Appendix~B of~\cite{islam2025anomaly}. 

We selected this dataset because it combines production-scale telemetry logs with manually verified anomaly labels, providing a realistic benchmark for anomaly detection in operational cloud environments. To the best of our knowledge, this is the first study to use the contextual structure encoded in the IBM Cloud Telemetry Dataset's feature schema to construct graph representations for anomaly detection in \LCSs. More importantly, the telemetry collected in this dataset reflects several characteristics commonly encountered in production cloud environments that complicate building an anomaly detection framework in \LCSs:

\begin{itemize}
\item[\textbf{C1}] \textbf{High-dimensional Telemetry}, resulting from the large number of monitored API activities and heterogeneous performance metrics.

\item[\textbf{C2}] \textbf{Spatio-temporal Dependencies}, arising from interactions among distributed cloud services and their deployment topology.

\item[\textbf{C3}] \textbf{Telemetry Sparsity}, where many API activities exhibit intermittent or low activity, resulting in sparse telemetry matrices and highly imbalanced temporal activity patterns (see Appendix~B in~\cite{islam2025anomaly}).
\end{itemize}

These characteristics motivate the design of \cloudens and provide the foundation for the comprehensive empirical study presented in this paper, which investigates anomaly detection performance under realistic production cloud conditions.

\subsection{Problem Formulation\label{sec:problem_statement}}

\noindent\textit{Time Series Notations.} 
Let $N$ denote the number of monitored API activities, $M$ denote the number of numerical performance metrics collected for each interaction (e.g., request count, average response time), and $T$ denote the total number of timestamps in the observation period.
For an API activity $e$ and numerical performance metric $p$, the temporal evolution of the metric is represented as a univariate time series
\begin{equation}
    \mathbf{x}_{e}^{p} = \left[ x_{e}^{p}(t_1),\, x_{e}^{p}(t_2),\, \ldots,\, x_{e}^{p}(t_T) \right] \in \mathbb{R}^{T},
    \label{eq:univariate_ts}
\end{equation}
where $x_{e}^{p}(t_i)$ denotes the observed value of performance metric $p$ at timestamp $t_i$. Consider a cloud system with $N$ monitored API activities, where $M$ numerical performance metrics are continuously collected from each interaction. The per-metric telemetry matrix $\mathbf{X}^{p}$ and the complete telemetry matrix $\mathbf{X}$ are defined as
\begin{equation}
    \mathbf{X}^{p} = 
    \begin{bmatrix} \mathbf{x}_{e_1}^{p} \\ \vdots \\ \mathbf{x}_{e_N}^{p} 
    \end{bmatrix} \in \mathbb{R}^{N \times T},
    \qquad
    \mathbf{X} = 
    \begin{bmatrix} \mathbf{X}^{1} \\ \vdots \\ \mathbf{X}^{M} 
    \end{bmatrix} \in \mathbb{R}^{(N \cdot M) \times T}.
    \label{eq:entire_performance_matrix}
\end{equation}
Each row of $\mathbf{X}$ represents the temporal evolution of a numerical performance metric associated with an API activity, while each column captures the operational state of the cloud monitoring system at a particular timestamp. 

\noindent\textit{Forecasting-based Anomaly Detection.} Given $\mathbf{X}$ representing telemetry monitoring data in \LCSs, our objective is to develop an automated multivariate time-series anomaly detection (MTSAD) solution that enables early identification and resolution of unexpected failures.
As recent advances in MTSAD are largely driven by forecasting-based 
methods~\cite{boniol2024divetimeseriesanomalydetection, Paparrizos2025}, we 
formulate our task as a forecasting problem, in which forecasting models are built to predict future telemetry records from historical observations, and significant forecasting errors indicate anomalous behavior. In this formulation, a row of $\mathbf{X}$ is also commonly referred to as an
\textit{input feature} in the machine learning community, where each 
feature represents a specific performance metric of an API endpoint in the context of telemetry monitoring in \LCSs.
Accordingly, we use the terms \textit{telemetry feature} and \textit{input
feature} interchangeably throughout the remainder of this paper. 

Consider a subset of telemetry features with $F$ features, where 
$1 \leq F \leq N \cdot M$, forming the matrix $\mathbf{X}' \in 
\mathbb{R}^{F \times T}$. In the special case where the subset consists only 
of features associated with a single performance metric $p$, we have $F = N$, and the resulting telemetry feature matrix reduces to $\mathbf{X}' = \mathbf{X}^p \in \mathbb{R}^{N \times T}$, as defined in Equation~\eqref{eq:entire_performance_matrix}. Given a sliding window of $w$ historical 
observations
\begin{equation}
    \mathbf{X}'(t-w+1), \mathbf{X}'(t-w+2), \ldots, \mathbf{X}'(t),
    \label{eq:sliding_window_subset}
\end{equation}
the objective is to predict the operational state $\hat{\mathbf{X}}'(t+1) \in \mathbb{R}^{F}$. The forecasting error vector is computed as
\begin{equation}
    \mathbf{E}(t+1) = \left| \hat{\mathbf{X'}}(t+1) - \mathbf{X'}(t+1) \right| \in \mathbb{R}^{F}.
    \label{eq:forecast_errors}
\end{equation}
Considering all timestamps with only $F$ telemetry features being used, the forecasting error 
vectors form a matrix
\begin{equation}
    \mathbf{E} = \left[ \mathbf{E}(1), \mathbf{E}(2), \ldots, \mathbf{E}(T) \right] 
    \in \mathbb{R}^{F \times T}.
    \label{eq:forecasting_error_matrix}
\end{equation}
An anomaly scoring function $g: \mathbb{R}^{F \times T} \rightarrow \mathbb{R}^{T}$ transforms 
the forecasting error matrix to an
anomaly score sequence
\begin{equation}
    \mathbf{s} = g\left( \mathbf{E} \right) = [ s(1),\, s(2),\, \ldots,\, s(T) ]\in \mathbb{R^T},
    \label{eq:anomaly_score_sequence}
\end{equation}
where higher values of $s(t)$ directly indicate a higher likelihood of anomalous system behavior. The resulting anomaly score
sequence is subsequently thresholded to identify candidate anomalous timestamps.

\noindent\textbf{Objective.} Given the telemetry matrix $\mathbf{X} \in \mathbb{R}^{(N \cdot M) \times T}$, 
which represents the operational behavior of monitored cloud services in \LCSs, our objective is to build an effective, 
forecasting-based anomaly detection solution which satisfies two key operational requirements: (i) \textit{early detection}, where anomalies are identified as soon as possible after their onset to enable timely intervention before cascading failures occur, and (ii) \textit{broad anomaly coverage}, which reflects the effectiveness of the monitoring system, while keeping false alarms under control to reduce operational overhead. Achieving this objective requires addressing the intrinsic characteristics of cloud telemetry logs, including high-dimensional feature spaces, spatio-temporal dependencies, and severe telemetry sparsity, corresponding to the three challenges C1, C2, and C3 introduced in Section~\ref{sec:telemetry_logs}.

\begin{figure*}[t]
  \centering
  \includegraphics[width=\linewidth]{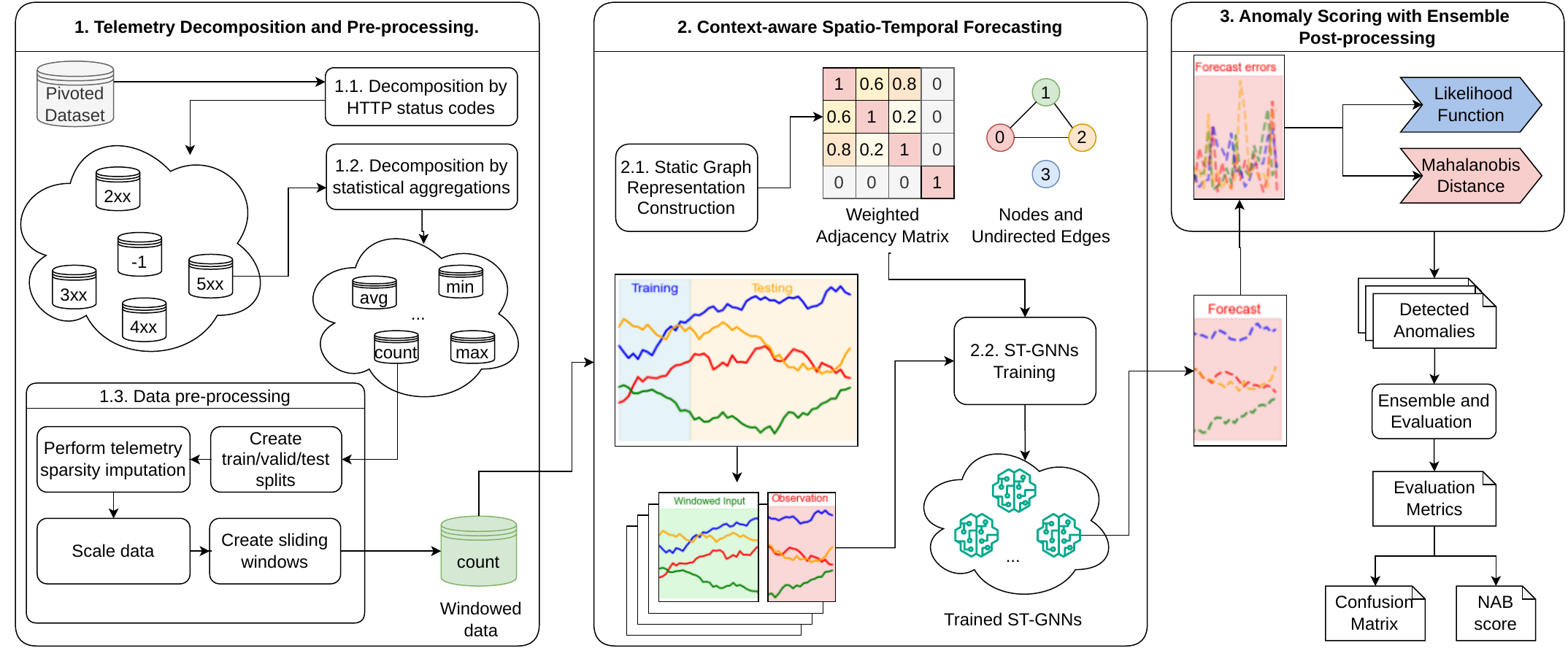}
\caption{Workflow of \cloudens. The pivoted telemetry dataset is partitioned into telemetry subsets based on HTTP status code groups (\texttt{4xx} and \texttt{5xx}) and aggregation functions (\texttt{count}, \texttt{avg}, \texttt{min}, and \texttt{max}). Each subset is processed independently by an ST-GNN, followed by anomaly scoring, thresholding, and subset-based ensemble decision to produce detected anomalies.}
  \label{fig:cloudens_architecture}
\vspace{-10pt}
\end{figure*}
\section{Related Work\label{sec:related_work}}

Anomaly detection is a fundamental capability of cloud monitoring because even short service disruptions can significantly affect application availability and operational cost. Cloud providers continuously collect telemetry data, including metrics, logs, and traces, to monitor service health and identify abnormal behaviors before they propagate across distributed services~\cite{Gatev2021LogMetricTrace,islam2021anomaly,islam2025anomaly,Chu2025_log_entity_graph,Hrusto2026_monitoring_data_in_cloud_system}. These telemetry records are typically represented as multivariate time series and often include contextual information such as service dependencies, deployment topology, and communication relationships. Recent surveys and benchmarking studies~\cite{boniol2024divetimeseriesanomalydetection,Paparrizos2025,Hrusto2026_monitoring_data_in_cloud_system,islam2026benchmarkinganomalydetectionheterogeneous} show that although numerous anomaly detection methods have been proposed, no single approach consistently performs well across different cloud environments because telemetry data differ substantially in dimensionality, sparsity, correlation structure, and operational characteristics.

Early anomaly detection methods mainly relied on statistical models and traditional machine learning techniques, including clustering, Principal Component Analysis (PCA), Isolation Forest, and One-Class SVM~\cite{Aggarwal2001,Schmidt2018Arima,HANYFAWZY20238,Qingfeng2018KNN,Traini_2023_k_mean,Kalaki2023PCA,Kardani2019Isolation,Shuaibo2024SVM}. These approaches are computationally efficient and relatively easy to deploy, but they generally assume static data distributions and therefore struggle to model the dynamic behavior of Large-scale Cloud Systems. Their performance further degrades when applied to high-dimensional and sparse telemetry commonly observed in production environments~\cite{thudumu2020survey}.

To better capture nonlinear temporal patterns, recent studies have increasingly adopted deep learning techniques such as Convolutional Neural Networks (CNNs), Recurrent Neural Networks (RNNs), autoencoders, and Transformer-based architectures~\cite{Kale_2022_CNN,Abdelsalam2019CNN,Du2017SystemLogDeepLearning,liu2019deepanomalydetectionpacketLSTM,islam2025anomaly,Hagemann2020Autoencoder,guo2020peekinsideclosedworld,liu2024itransformerinvertedtransformerseffective}. These methods generally achieve higher detection accuracy by learning complex temporal dependencies directly from telemetry data. Nevertheless, most of them model cloud telemetry as independent feature sequences and therefore do not explicitly exploit the structural relationships among distributed cloud services.

Graph Neural Networks (GNNs) have recently emerged as an effective solution for modeling relational dependencies in multivariate time series~\cite{CHEN2023TraceGraGNN,Zeng2024,jin2024surveygraphneuralnetworks}. Graph Convolutional Networks (GCNs)~\cite{defferrard2017convolutionalneuralnetworksgraphs} and Graph Attention Networks (GATs)~\cite{Petar2018graphattentionnetworks} enable anomaly detection models to leverage service dependencies by representing telemetry features as graph nodes connected through operational relationships. Building on these developments, Spatio-Temporal Graph Neural Networks (ST-GNNs) jointly model temporal dynamics and graph topology, making them particularly suitable for microservice-based cloud environments~\cite{Golovkina2024_gnn_micro_service,Kevin2025_graph_cloud_microservice}. Existing studies have demonstrated promising results for metric prediction and anomaly detection, but most evaluations rely on simulated workloads or focus primarily on prediction accuracy.

Despite these advances, several challenges remain insufficiently addressed in production cloud monitoring, including extreme dimensionality, heterogeneous metrics, complex inter-service dependencies, and severe sparsity, with limited systematic investigation of context-aware modeling, anomaly scoring, and sparsity imputation on real-world telemetry. These challenges are clearly reflected in the IBM Cloud Telemetry Dataset~\cite{islam2025anomaly}. Based on a deeper analysis of this dataset, we propose an anomaly detection framework tailored for \LCSs and conduct an empirical study investigating how post-processing techniques, such as scoring strategies and ensembling, affect the trade-off between anomaly coverage and false alarms in \LCS monitoring.

\section{The proposed framework: \cloudens \label{sec:cloudens_framework}}

This section presents \cloudens, an operational context-aware anomaly detection framework for cloud telemetry collected from \LCSs that jointly leverages contextual information and numerical performance metrics through telemetry feature decomposition, context-aware graph modeling, spatio-temporal forecasting, anomaly scoring, and ensemble post-processing. Section~\ref{sec:overview} provides an overview of the framework, while the subsequent subsections describe each stage in detail.

\subsection{Overview\label{sec:overview}}
Fig.~\ref{fig:cloudens_architecture} presents an overview of the proposed \cloudens framework, where the design aims to address the key challenges of anomaly detection in \LCSs, as outlined in Section~\ref{sec:telemetry_logs}. 
It follows the common framework used in time-series anomaly detection~\cite{boniol2024divetimeseriesanomalydetection,Paparrizos2025}, which includes the stages of \textit{pre-processing}, \textit{detection}, \textit{anomaly scoring}, and \textit{post-processing}. As shown in Fig.~\ref{fig:cloudens_architecture},  \cloudens organizes these stages into three main modules, combining the latter two stages into a single component. They are summarized as follows:

\begin{itemize}

    \item[\textbf{(1)}] \textbf{Telemetry Decomposition and Pre-processing.} To address the high dimensionality of cloud telemetry (\textbf{C1}), \cloudens partitions telemetry data into semantically coherent subsets based on the contextual information encoded in API activity naming templates (Listing~\ref{lst:feature_template}). This decomposition enables specialized detection models to model complementary aspects of cloud service behavior.

    \item[\textbf{(2)}] \textbf{Context-aware Spatio-Temporal Forecasting.} This stage learns the normal operational behavior of each telemetry feature subset through a forecasting-based approach. To jointly capture temporal dynamics and service dependencies (\textbf{C2}), \cloudens constructs context-aware graphs from contextual information and employs Spatio-Temporal Graph Neural Networks (ST-GNNs) as the forecasting backbone.
    
    \item [\textbf{(3)}] \textbf{Anomaly Scoring with Ensemble Post-processing.} Forecasting errors are transformed into anomaly scores using statistical scoring strategies and thresholded to identify anomalous timestamps. The resulting anomaly decisions from selected telemetry subsets are then combined using a one-vote ensemble strategy to improve anomaly coverage and produce the final detection results.

\end{itemize}
\subsection{Telemetry Decomposition and Pre-processing\label{sec:feature_split}}

In \LCSs, telemetry logs collected from distributed microservices are often high-dimensional and heterogeneous (refer to \textbf{C1} in Section~\ref{sec:telemetry_logs}). In such settings, anomalies become difficult to distinguish from normal behavior when analyzed in the full feature space. Furthermore, irrelevant or noisy attributes can obscure meaningful deviations, as abnormal patterns are often manifested within only a small subset of features rather than across the entire dimensionality~\cite{Zimek2012_high_dimenstional}. In addition, training anomaly detection models on the full monitored data increases computational cost and hinders fault localization. To deal with these issues, our approach leverages contextual information in telemetry logs to divide them into smaller, more manageable telemetry subsets. This module groups raw telemetry logs into semantically meaningful subsets, enabling more targeted and cost-effective model training.

The contextual attributes associated with each API activity, including deployment location, communication role, HTTP status code, endpoint identifier, and aggregation type, provide semantic descriptions of cloud service operations. Our framework exploits these attributes to partition telemetry features into semantically coherent subsets. Using the IBM Cloud Telemetry Dataset as an example, we first group telemetry features according to HTTP status code categories (e.g., \texttt{4xx} and \texttt{5xx}) to separate client-side and server-side behaviors. Each status-code group is then further divided by aggregation type (e.g., count, average, minimum, and maximum), hereafter referred to as \texttt{count}, \texttt{avg}, \texttt{min}, and \texttt{max}. This decomposition produces multiple operationally meaningful telemetry subsets, allowing anomaly detection models to focus on homogeneous behaviors while reducing the impact of irrelevant or noisy features. Fig.~\ref{fig:cloudens_architecture}(1) illustrates this decomposition process.

This decomposition offers several key advantages. First, it reduces input dimensionality while preserving semantic structure, which lowers model complexity and improves training efficiency. Second, it enables fine-grained anomaly detection, as certain anomalies may manifest prominently in one subset (e.g., a spike in \texttt{5xx count}) but remain undetectable in others. 
Overall, we perform this decomposition strategy to transform raw, high-dimensional, heterogeneous telemetry logs into multiple, small, and homogeneous telemetry subsets, forming the foundation of the ensemble-based architecture used in \cloudens. Each subset is subsequently preprocessed and independently processed by the context-aware forecasting stage described in the next subsection.

\subsection{Context-aware Spatio-Temporal Forecasting\label{sec:stgnn_forecasting}}

\begin{figure}[t]
  \centering
  \includegraphics[width=\linewidth]{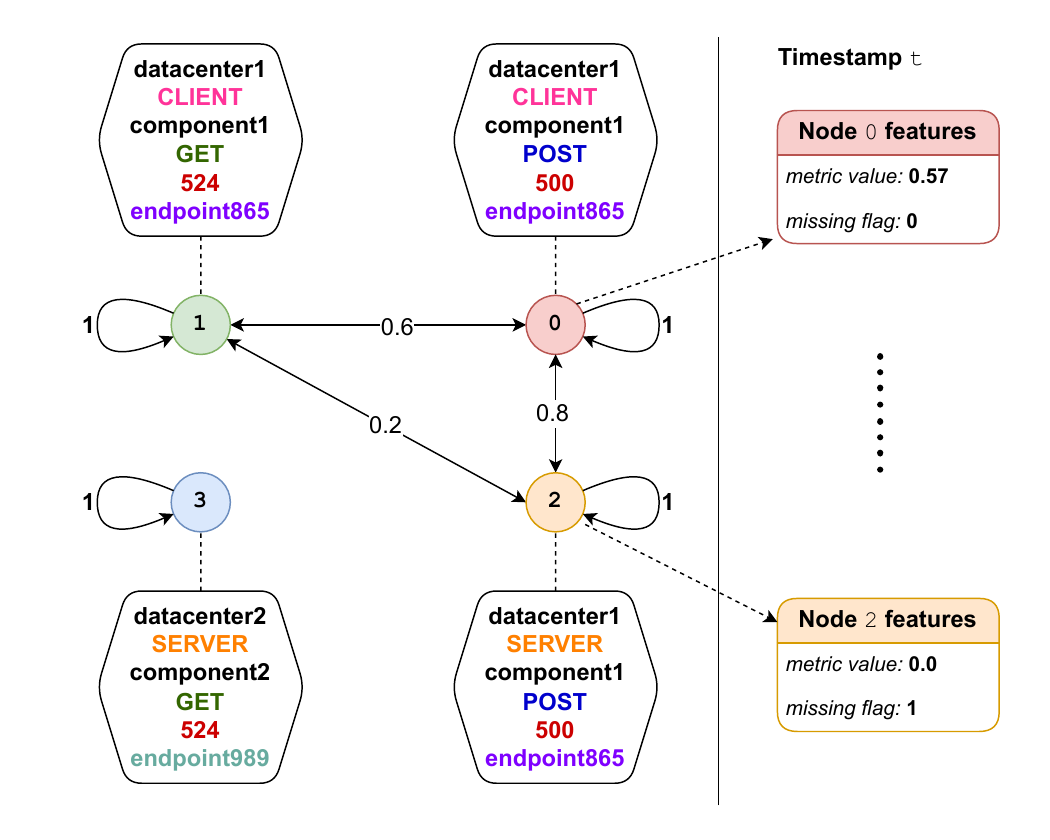}
  \caption{Context-aware graph: nodes are API activities; edge weights reflect the strength of shared operational-context attributes (e.g., component, endpoint). 
  }
  \label{fig:graph}
  \vspace{-15pt}
\end{figure}

Recent advances in MTSAD have introduced a wide range of learning paradigms, including reconstruction-, forecasting-, and representation-based approaches~\cite{Paparrizos2025,boniol2024divetimeseriesanomalydetection, Hrusto2026_monitoring_data_in_cloud_system}. Among these, forecasting-based methods have become one of the most effective paradigms because they learn normal system behavior from historical observations without relying on anomaly labels, making them particularly suitable for real-world applications, where labeled anomalies are scarce.
Although many forecasting-based approaches have been proposed in the literature across different contexts~\cite{boniol2024divetimeseriesanomalydetection, Hrusto2026_monitoring_data_in_cloud_system,corradini2025systematicliteraturereviewspatiotemporal}, our study employs Spatio-Temporal Graph Neural Networks (ST-GNNs) as the main forecasting backbone because they jointly model temporal dynamics and dependencies among correlated variables~\cite{jin2024surveygraphneuralnetworks}. This capability is particularly well suited to work with telemetry monitoring data from \LCSs, where API activities exhibit complex relationships induced by deployment topology and service dependency. 
The remainder of this section presents the concept of applying ST-GNNs to a single telemetry 
subset produced by the previous module.

\textbf{Definition 1.} \textit{Context-aware graph $\mathcal{G}$}: The 
operational context of telemetry monitoring data is described as 
$\mathcal{G} = (\mathcal{V}, \mathcal{E})$, where $\mathcal{V} = \{v_1, v_2, 
\ldots, v_N\}$ is the set of $N$ nodes representing $N$ API activities. $\mathcal{E}$ is the set of edges, 
which reflects the operational-context relationships between API activities (e.g., shared deployment location, service ownership, communication 
role, HTTP method, or endpoint identifier). All of the relationship 
information is stored in the weighted adjacency matrix $\mathbf{A} \in 
\mathbb{R}^{N \times N}$, where rows and columns are indexed by telemetry 
features, and the value of each entry indicates the degree of contextual 
correlation between the corresponding features. The entry value is zero if 
the two features share no common operational-context attributes, and positive otherwise, with larger values reflecting stronger operational relationships; self-loops are set to one.

For each telemetry feature subset, \cloudens constructs a context-aware graph $\mathcal{G}$ following Definition 1.
Fig.~\ref{fig:graph} illustrates an example of the proposed graph representation using telemetry features from the IBM Cloud Telemetry Dataset. In this example, four representative nodes correspond to telemetry features identified by contextual attributes, including deployment location, communication role, host, HTTP method, status code, endpoint identifier, and aggregation type. An undirected edge is established when two nodes share common operational attributes, indicating that the corresponding API activities are likely to exhibit correlated behavior. For example, nodes \texttt{0}, \texttt{1}, and \texttt{2} are connected because they belong to the same deployment location (\texttt{datacenter1}), service component (\texttt{component1}), and endpoint (\texttt{endpoint865}), whereas node \texttt{3} remains isolated because it represents a different API activity.
Furthermore, the edge between nodes sharing the same deployment location, 
host, HTTP method, and endpoint (nodes $0$ and $2$) receives a larger weight (0.8) than the edge between nodes differing in communication role or HTTP method (nodes $1$ and $2$), which receives a weight of 0.2, as shown in Fig.~\ref{fig:graph}. 

Our purpose of constructing a context-aware graph is to enable the 
forecasting model to exploit operational dependencies among distributed cloud services (Challenge C2, 
Section~\ref{sec:telemetry_logs}), rather than treating each API activity as an independent telemetry feature. This is particularly important for detecting 
localized anomalies, since a failure in one service component often 
propagates to correlated API activities, and jointly modeling these relationships alongside temporal patterns in the time series can improve detection performance.

\subsubsection{Node Feature Engineering\label{sec:node_features}}

Each node in the context-aware graph corresponds to a telemetry feature, whose primary attribute is the associated numerical performance metric. Since cloud telemetry is inherently sparse, with many API activities remaining inactive during substantial portions of the monitoring period, \cloudens optionally augments each node with a binary telemetry sparsity indicator. This auxiliary node feature enables the forecasting model to distinguish genuine operational behavior from values introduced through telemetry sparsity imputation.

Consider $\mathbf{X}'=\mathbf{X}^p \in \mathbb{R}^{N\times T}$ to represent the telemetry matrix of a single feature subset of the performance metric $p$ according to Equation~\ref{eq:entire_performance_matrix}, where $\mathbf{X}'_e(t)$ represents the performance metric of node $e$ at timestamp $t$. The corresponding sparsity indicator matrix $\mathbf{M} \in \{0,1\}^{N\times T}$ is defined as
\begin{equation}
\mathbf{M}_e(t)=
\begin{cases}
1, & \text{if } \mathbf{X}'_e(t) \text{ exhibits no activity},\\
0, & \text{otherwise}.
\end{cases}
\label{eq:missing_indicator}
\end{equation}
The node feature representation is then extended to
\begin{equation}
\mathbf{\tilde{X}}'\in\mathbb{R}^{N \times2\times T},
\label{eq:extended_node_feature}
\end{equation}
where the feature vector of node $e$ at $t$ is given by
\begin{equation}
\mathbf{\tilde{X}}'_e(t)=
\left[
\mathbf{X}'_e(t),
\mathbf{M}_e(t)
\right]
\in\mathbb{R}^{2}.
\label{eq:node_feature}
\end{equation}
Fig.~\ref{fig:graph} illustrates this representation. For example, node \texttt{0} is represented by $[0.57,\,0]$, indicating a normalized telemetry value of $0.57$ with a valid observation, whereas node \texttt{2} indicates a zero-imputed telemetry value together with a sparsity flag. By explicitly encoding sparsity, the forecasting model can distinguish sparse telemetry entries from normal operational behavior, improving detection capability under severe telemetry sparsity (Challenge~\textbf{C3}, 
Section~\ref{sec:telemetry_logs}).

\subsubsection{Forecasting-based models with ST-GNNs}
The engineered node feature matrix $\mathbf{\tilde{X}}'$, together with the context-aware graph $\mathcal{G}$, is provided as input to build the ST-GNN forecasting model. Given a sequence of historical graph snapshots, the model learns the normal operational behavior of cloud services by jointly exploiting temporal dynamics and operational-context dependencies to predict the telemetry monitored values at the next timestamp. Formally, the forecasting process is defined as
\begin{equation}
\hat{\mathbf{X}}'(t+1)
=
f_{\mathrm{ST\mbox{-}GNN}}
\!\left(
\mathbf{\tilde{X}}'(t-w+1:t),\,\mathcal{G}
\right),
\label{eq:stgnn_forecasting}
\end{equation}
where $f_{\mathrm{ST\mbox{-}GNN}}(\cdot)$ denotes the forecasting model, $\mathbf{\tilde{X}}'(t-w+1:t)$ represents the sequence of engineered node features over a sliding window of length $w$, and $\mathcal{G}$ is the context-aware graph constructed according to Definition~1. The predicted value vector $\hat{\mathbf{X}}'(t+1)$ is subsequently compared with the ground-truth observation $\mathbf{X}'(t+1)$ to compute the forecasting error vector, as defined in Equation~\ref{eq:forecast_errors}.

\subsection{Anomaly Scoring with Ensemble Post-processing\label{sec:anomaly_scorer}}

The forecasting models produce a prediction error sequence $\mathbf{E}(t)$ for each telemetry feature subset, which is subsequently transformed into anomaly scores before anomaly decisions are made. The choice of anomaly scoring strategy directly influences detection performance because different telemetry subsets exhibit distinct statistical characteristics and therefore respond differently to the same scoring function~\cite{Malhotra2015LongST,Kejiang17,hewamalage2022forecastevaluationdatascientists}. To account for this variability, \cloudens evaluates two representative anomaly scoring strategies that provide complementary perspectives on anomalous behavior. They are briefly described as follows.

\begin{itemize}

\item[\textbf{(1)}] \textbf{Likelihood Function (LF)}~\cite{Lavin_2015,AHMAD2017134StreamingData}. 
The Likelihood Function estimates the anomaly likelihood by modeling the 
temporal distribution of forecasting errors using two rolling statistical 
windows of sizes $W$ (long window) and $W'$ (short window), which capture the historical and recent error distributions, respectively. The resulting likelihood at timestamp $t$, regarded as $s(t) \in (0,1)$, is obtained by comparing these two distributions, and the timestamp is classified as anomalous when $s(t) \geq L_t$, where $L_t$ is a user-defined threshold. The parameters $W$, $W'$, and $L_t$ control the sensitivity of detection performance and are optimized empirically.

\item[\textbf{(2)}] \textbf{Mahalanobis Distance (MD)}~\cite{MahalanobisOriginal2018,Christophe2018MahalanobisDistance,PANG2023MahalanobisTemporal}. 
The Mahalanobis Distance measures how far the multivariate forecasting error deviates from the historical error distribution while accounting for correlations among dimensions. It is computed as
\begin{equation}
d_M(\mathbf{E}(t))
=
\sqrt{
(\mathbf{E}(t)-\boldsymbol{\mu})^\top
\mathbf{\Sigma}^{-1}
(\mathbf{E}(t)-\boldsymbol{\mu})
} \in \mathbb{R},
\label{eq:mahalanobis}
\end{equation}
where $\mathbf{E}(t)$ denotes the forecasting error vector at timestamp $t$, $\boldsymbol{\mu}$ is the mean forecasting error, and $\mathbf{\Sigma}^{-1}$ is the inverse covariance matrix estimated from historical forecasting errors. 
Over the observation period, the Mahalanobis Distance values form the 
sequence $D_M \in \mathbb{R}^{T}$, min-max normalized to $[0,1]$ to 
obtain the anomaly score sequence $\mathbf{s}$ as defined in Equation~\ref{eq:anomaly_score_sequence}. A timestamp $t$ is classified as anomalous if $s(t)$ ranks among the top $(100-\epsilon)\%$ of $\mathbf{s}$, where $\epsilon \in (0, 100)$.

\end{itemize}

Since different telemetry subsets characterize complementary operational aspects of cloud services, they often reveal different anomaly patterns. Rather than relying on a single detector, \cloudens aggregates the anomaly decisions from selected telemetry subsets using a one-vote ensemble strategy, where a timestamp is classified as anomalous if it is detected by at least one detector. This ensemble post-processing improves anomaly coverage, though typically at 
the cost of an increased false alarm rate. The influence of different scoring strategies and ensemble configurations is investigated in Section~\ref{sec:evaluation}.

\section{Evaluation \label{sec:evaluation}}

Our evaluation is organized around the following three evaluation questions, each investigating a key design aspect of \cloudens and providing practical guidance for developing reliable anomaly detection frameworks for \LCSs.

\begin{itemize}[leftmargin=1cm]
\item [\textit{\textbf{EQ$_1$}}:] \rqfirst
\item [\textit{\textbf{EQ$_2$}}:] \rqsecond 
\item [\textit{\textbf{EQ$_3$}}:] \rqthird
\end{itemize}

\subsection{Datasets and Forecasting Models}

\subsubsection{Datasets} The evaluation is conducted on the publicly available IBM Cloud Console telemetry dataset introduced by Islam \emph{et al.}~\cite{islam_2024_14062900,islam2025anomaly}. As described in Section~\ref{sec:working_dataset}, the dataset contains telemetry collected from a production cloud environment over approximately 4.5 months, comprising 39,365 timestamps and 117,448 telemetry features.
From a cloud monitoring perspective, we focus on telemetry features associated with HTTP \texttt{4xx} and \texttt{5xx} status codes because they directly capture client- and server-side failures, making them strong indicators of abnormal behavior in {\LCSs}. Each status-code category is partitioned into four statistical aggregation subsets (\texttt{count}, \texttt{avg}, \texttt{min}, and \texttt{max}),
resulting in eight telemetry subsets for evaluation. Each \texttt{5xx} subset contains 2,406 telemetry features with 99.02\% sparsity, whereas each \texttt{4xx} subset contains 4,085 features with 94.74\% sparsity.

\subsubsection{Forecasting Models}
Following the problem formulation in Section~\ref{sec:problem_statement}, anomaly detection is formulated as a one-step-ahead forecasting task~\cite{taieb2011reviewcomparisonstrategiesmultistep}, where anomalies are identified from deviations between predicted and observed telemetry. This formulation is well suited to cloud monitoring because streaming telemetry can be continuously evaluated against expected system behavior inferred from recent historical observations. Although numerous forecasting models have been proposed for multivariate time-series anomaly detection, their effectiveness varies considerably across datasets, evaluation metrics, and application domains~\cite{Wenig2024,Paparrizos2025}. Therefore, rather than proposing a new forecasting architecture, this work focuses on understanding the characteristics of cloud telemetry and designing an anomaly detection framework tailored to \LCSs. To evaluate the effectiveness of context-aware forecasting, we compare the following forecasting models.

\begin{itemize}

\item \textbf{GRU}~\cite{cho2014propertiesneuralmachinetranslation}: The Gated Recurrent Unit (GRU) is a recurrent neural network that models temporal dependencies through update and reset gating mechanisms, enabling it to capture long- and short-term sequential patterns while mitigating the vanishing gradient problem.

\item \textbf{A3T-GCN}~\cite{zhu2020a3tgcnattentiontemporalgraph}: A3T-GCN is an attention-based Spatio-Temporal Graph Neural Network that combines temporal graph convolutions with an attention mechanism to jointly model temporal dynamics and structural dependencies among telemetry features. The weighted adjacency matrix is derived from the operational-context attributes (see Section~\ref{sec:telemetry_logs}).
\end{itemize}

Although A3T-GCN serves as the forecasting backbone of \cloudens, the proposed framework is model-agnostic and can readily incorporate other ST-GNN architectures according to empirical evaluation and application-specific requirements.

\subsection{Experimental settings}

\subsubsection{Hardware and Environment}
All experiments are conducted on a high-performance computing (HPC) cluster 
equipped with an NVIDIA A100 GPU (80 GB), running Python 3.11.6, PyTorch 
2.12.1, and CUDA 12.6. ST-GNN models are implemented using PyTorch Geometric 
Temporal~\cite{pytorch_geometric_temporal}. Our reproducible package is publicly available at~\cite{replicationPackage}.

\subsubsection{Dataset preparation}
\begin{table}[H]
  \caption{Summary of Ground-Truth Anomalies by Source in the testing period used in \cite{islam2025anomaly}, with an illustration in Fig.~\ref{fig:5xx_count_mts}.}
  \label{tab:anomaly_summary}
  \begin{center}
  \begin{tabular}%
  {p{3.7cm}cl}
    \toprule
    \text{Anomaly Source} & \text{Count} & \text{Anomaly IDs} \\
    \midrule
    \colorbox{OliveGreen!20}{Issue Tracker}     & 3  & 3, 12, 13 \\
    \colorbox{orange!20}{Instant Messenger} & 9  & 0, 5, 6, 7, 8, 9, 14, 16, 17 \\
    \colorbox{violet!20}{Test Log}          & 7  & 1, 2, 4, 10, 11, 15, 18 \\
    \hline
    \bottomrule
  \end{tabular}
  \end{center}
  \vspace{-10pt}
\end{table}

\begin{table}[!b]
  \caption{Cost profiles of NAB Score~\cite{Lavin_2015}.}
  \label{tab:cost_profiles}
  \centering
  \resizebox{\columnwidth}{!}{
  \begin{threeparttable}
  \begin{tabular}{p{3cm}cccc}
    \toprule
    \text{Profile} & \multirowcell{2}{\text{TP}\\ \text{Weight}} & \multirowcell{2}{\text{FN}\\ \text{Weight}} & \multirowcell{2}{\text{FP}\\ \text{Weight}} & \multirowcell{2}{\text{TN}\\ \text{Weight}}\\
    &&&& \\
    \midrule
    Standard     & 1.00  & 1.00 & 0.11 & 1.00 \\
    Reward Low FN\tnote{*} &  1.00  & 2.00 & 0.11 & 1.00 \\
    \hline
    \bottomrule
  \end{tabular}
      \begin{tablenotes}
        \item[*] Hereafter, the \textit{Reward Low FN} profile is abbreviated as Low FN.
    \end{tablenotes}
  \end{threeparttable}
  }
\end{table}

Following the experimental protocol of Islam \emph{et al.}~\cite{islam2025anomaly}, we use five weeks of telemetry (January~26,~2024 to February~29,~2024) for training and the subsequent three months (March~1,~2024 to May~31,~2024) for testing. We reserve 30\% of the training data as a validation set to monitor training convergence and tune the hyperparameters of the scoring strategies.
Ground-truth anomalies are obtained from three independent operational sources: the \textit{Issue Tracker}, \textit{Test Log}, and \textit{Instant Messenger}. There are 25 annotated anomaly windows, of which 19 occur during the testing period. 
A summary of ground-truth anomalies in the testing period is provided in Table~\ref{tab:anomaly_summary}, where each anomaly is assigned a unique identifier that is used consistently throughout the paper. Note that each ground-truth anomaly in this table is represented as a sequence of contiguous timestamps rather than a single point; the 19 ground-truth anomalies cover 967 of the 26,488 testing-period timestamps, corresponding to an anomaly ratio of approximately 3.65\%.

Following the feature decomposition strategy described in Section~\ref{sec:feature_split}, the telemetry data are partitioned into eight independent telemetry subsets according to two HTTP status code categories (\texttt{4xx} and \texttt{5xx}) and four statistical aggregation types (\texttt{count}, \texttt{avg}, \texttt{min}, and \texttt{max}). The resulting subsets, namely \texttt{5xx count}, \texttt{5xx avg}, \texttt{5xx min}, \texttt{5xx max}, \texttt{4xx count}, \texttt{4xx avg}, \texttt{4xx min}, and \texttt{4xx max}, are evaluated independently to investigate the contribution of different telemetry feature groups to anomaly detection performance.

To address the sparsity of cloud telemetry, all subsets are evaluated with three imputation strategies: zero, mean, and median. Before training, each feature subset is normalized using min--max scaling based on the training data and transformed into input-output pairs using a sliding window of length $w$. Following common practice in forecasting-based anomaly detection, windows containing labeled anomalies are excluded from the training set to ensure that the forecasting models learn only normal operational behavior. The validation set is used for model selection and hyperparameter tuning, while the testing set is reserved exclusively for evaluation.

\subsubsection{Evaluation Metrics\label{sec:evaluation_metrics}}

To assess anomaly detection performance in the context of \LCSs, we consider common practices in prior work \cite{Sondre2023_ts_evaluation_metric,Paparrizos2025} and employ the Numenta Anomaly Benchmark (NAB) score~\cite{Lavin_2015} as the primary evaluation metric.

We first report the confusion matrix, including True Positives (TP), True Negatives (TN), False Positives (FP), and False Negatives (FN), to summarize the \textit{point-based} detection results. In the context of anomaly detection, the positive class represents anomalies, and the negative class denotes normal behavior. Although \textit{point-based} metrics derived from the confusion matrix, including Precision, Recall, F$_1$-score, and AUC, are widely used in anomaly detection research~\cite{Paparrizos2025}, they are less suitable for evaluating anomaly detection in \LCSs because they assess each timestamp independently. In practice, detecting at least one point within an anomaly window is often sufficient to identify an anomalous event and trigger corrective actions~\cite{AHMAD2017134StreamingData}.
Compared with traditional point-based metrics, the NAB score, a \textit{window-based} evaluation metric, provides a more realistic assessment for cloud service monitoring because it rewards early anomaly detection while penalizing delayed detections, missed anomalies, and false alarms.

The NAB score rewards early detection while penalizing late or missed detections via a scaled sigmoid function. Any alert within an anomaly window is treated as a single TP, with its contribution determined by a scaled sigmoid curve: detections at the start of the window receive high scores, while detections near the end yield lower scores. Additional alerts within the same window are ignored. Missing an entire anomalous window results in one FN, and FPs decrease the computed score. We evaluate anomaly detection performance using the Standard and Reward Low FN NAB cost profiles~\cite{Lavin_2015} (see Table~\ref{tab:cost_profiles}). The Standard profile balances rewards for early detection against penalties for FPs and FNs. The Reward Low FN profile applies a stronger penalty on false negatives, which aligns with the context of \LCSs where missing anomalies is considered critical. Using both NAB profiles provides DevOps teams with deeper insights into the trade-offs between early detection, missed anomalies, and false alarms, enabling more informed selection of anomaly detection configurations for different operational scenarios.

\subsubsection{Hyper-parameter Settings}
\paragraph{Forecasting Models}
Both the GRU baseline and the A3T-GCN forecasting model, applied independently to each subset, use the same empirically selected hyperparameters: a sliding window of 6 timestamps, a one-step forecasting horizon, 32 hidden channels, a batch size of 32, the Adam optimizer ($lr = 0.001$), and the MSE loss function. This configuration is used for \textbf{EQ$_1$} and \textbf{EQ$_2$}, while \textbf{EQ$_3$} evaluates sliding window sizes of $\{6,12,18,24,30\}$. Although A3T-GCN serves as the forecasting backbone of \cloudens, the framework is model-agnostic and can readily incorporate other ST-GNN architectures depending on the application requirements and the characteristics of the telemetry data.

\begin{table}[t]
\centering
\caption{Optimized imputation and anomaly scoring hyperparameters selected for each telemetry feature subset.}
\label{tab:scoring_hyperparameters}
\resizebox{\columnwidth}{!}{
\begin{threeparttable}
\begin{tabular}{lp{2cm}p{2cm}p{1.5cm}}
\toprule
\text{Subset} &
\text{\multirowcell{2}[0pt][l]{Sparsity\\Imputation}} &
\text{\multirowcell{2}[0pt][l]{Likelihood\\Function (LF)\tnote{a}}} &
\text{\multirowcell{2}[0pt][l]{Mahalanobis\\Distance (MD)}} \\
&&&\\
\midrule
$\texttt{4xx avg}$ & mean & $L_t=0.999$ & $\epsilon=99.6$ \\
$\texttt{4xx count}$ & mean & $L_t=0.9997$ & $\epsilon=99.5$ \\
$\texttt{4xx max}$ & median & $L_t=0.999$ & $\epsilon=99.6$ \\
$\texttt{4xx min}$ & zero & $L_t=0.998$ & $\epsilon=99.9$ \\
$\texttt{5xx avg}$ & mean & $L_t=0.999$ & $\epsilon=99.8$ \\
$\texttt{5xx count}$ & zero & $L_t=0.99975$ & $\epsilon=99.8$ \\
$\texttt{5xx max}$ & median & $L_t=0.998$ & $\epsilon=99.6$ \\
$\texttt{5xx min}$ & zero & $L_t=0.9998$ & $\epsilon=99.7$ \\

\hline
\bottomrule
\end{tabular}
\begin{tablenotes}
    \item[a] $W=30$ and $W'=2$ are fixed for all subsets.
\end{tablenotes}
\vspace{-15pt}
\end{threeparttable}
}
\end{table}

\input{tables/acm_table_1_graph_improvement}

\paragraph{Anomaly Scoring}
Following Section~\ref{sec:anomaly_scorer}, we evaluate two anomaly scoring strategies: the Likelihood Function (LF) and Mahalanobis Distance (MD). Both strategies transform forecasting errors into anomaly score sequences,  which are subsequently thresholded to classify anomalous and normal timestamps. For brevity, they are referred to as \textbf{LF} and \textbf{MD}, respectively, throughout the remainder of this paper. 
The hyperparameters of the two scoring strategies were determined through a grid search on the validation split. Specifically, for the LF scoring strategy, we evaluated the long window $W \in \{20,25,30,35,40,50\}$, the short window $W' \in \{1,2,3,4,5\}$, and threshold $L_t \in \{0.998, 0.9985, 0.999, 0.9997, 0.99975, 0.9998\}$. For MD scoring strategy, the threshold was varied over $\epsilon \in \{99.5,99.6,99.7,99.8,99.9\}$. The configuration achieving the highest NAB score under the Low FN profile is considered the optimal choice, which is reported in Table~\ref{tab:scoring_hyperparameters} for each subset. After tuning the scoring hyperparameters, the forecasting models are retrained on the combined training and validation splits before final evaluation on the testing period.

\subsection{Experimental results}

\begin{figure*}

    \begin{minipage}[t]{\textwidth}
        \includegraphics[width=\textwidth]{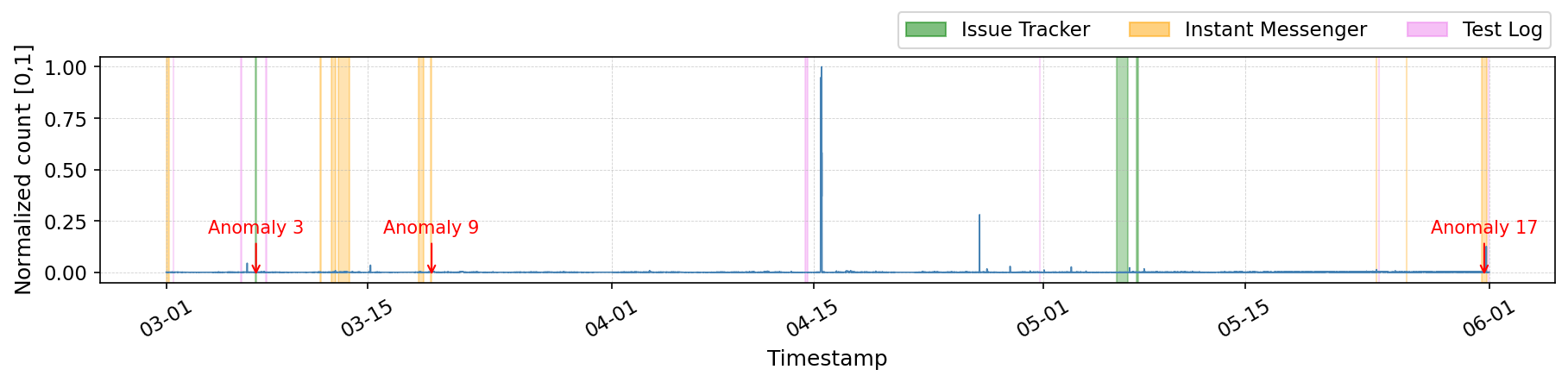}
        \caption{Sum of count of \texttt{5xx count} subset as MTS and several anomalies detected by \cloudens.}
        \label{fig:5xx_count_mts}
    \end{minipage}

    \begin{minipage}[t]{\textwidth}
        \includegraphics[width=\textwidth]{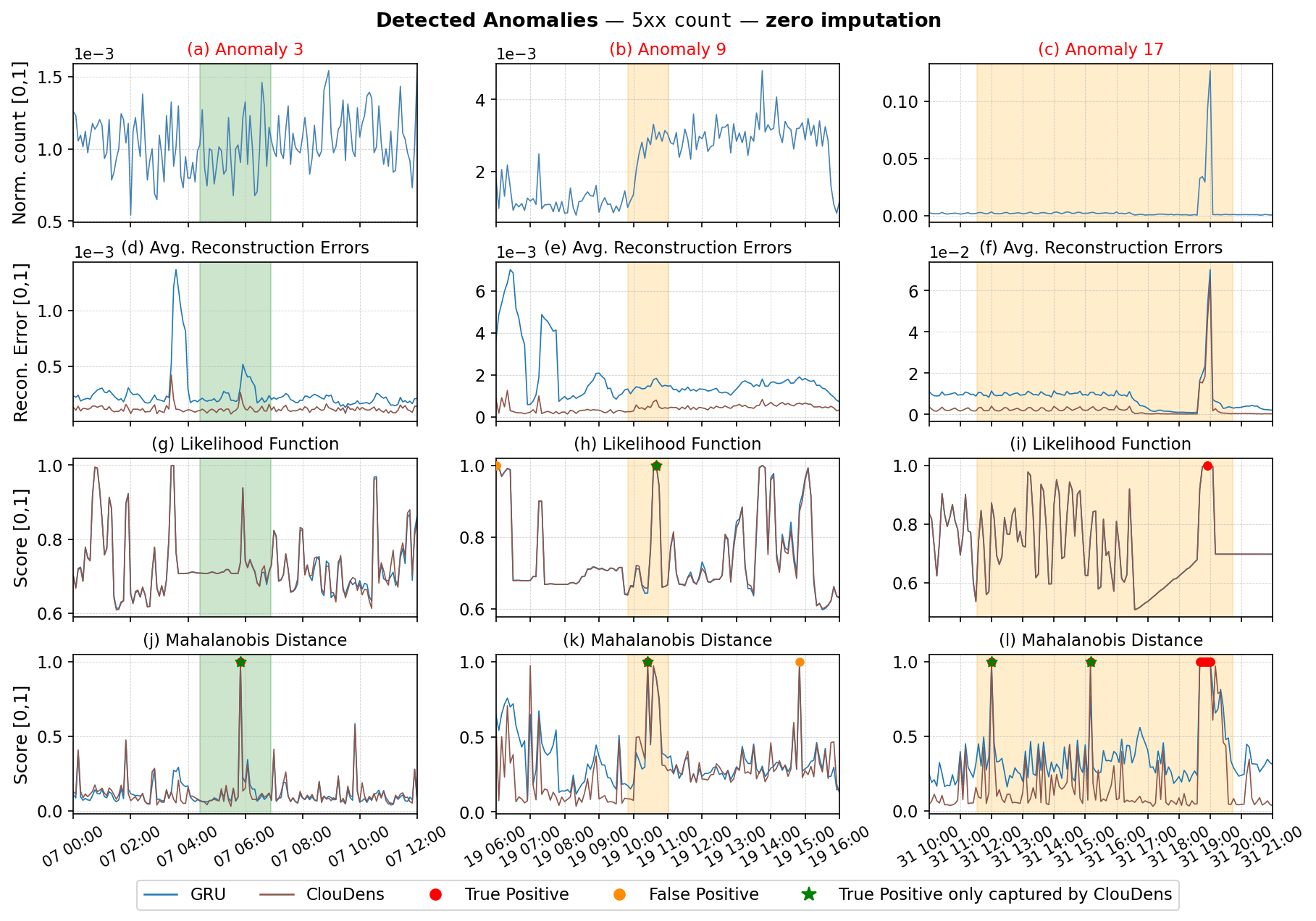}
        \caption{An illustration of detected anomalies from GRU and \cloudens.}
        \label{fig:5xx_count_zoom_in}
        \vspace{-10pt}
    \end{minipage}
\end{figure*}

\mysubsubsection{\textit{\textbf{EQ$_1$}}}{\rqfirst}
Table~\ref{tab:graph_improve} presents a performance comparison of \cloudens and GRU on the \texttt{5xx count} feature subset with respect to the ground-truth anomalies summarized in Table~\ref{tab:anomaly_summary}. The results are reported using the optimal thresholds selected for each scoring strategy, as reported in Table~\ref{tab:scoring_hyperparameters}. Under the LF scoring strategy, GRU detects 6 TPs, covering 5 ground-truth anomalies (Anomaly 12 in the \textit{Issue Tracker} category and Anomaly 6, 7, 8, and 17 in the \textit{Instant Messenger} category). GRU detects 53 FPs and 961 FNs, yielding NAB scores of 6.58 under Standard profile and 13.16 under Low FN profile. In contrast, \cloudens improves detection performance by identifying 7 TPs with fewer FPs (52) and FNs (960), increasing the NAB scores to 11.38 and 18.11 for Standard and Low FN profiles, respectively. The result of  detected ground-truth anomalies also shows that \cloudens successfully detects Anomaly~9 in the \textit{Instant Messenger} category, which GRU fails to capture.

Under the MD scoring strategy, GRU detects 13 TPs and produces 40 FPs, resulting in NAB scores of 5.89 and 10.95 for Standard and Low FN NAB cost profiles, respectively. \cloudens again outperforms GRU by identifying 16 TPs while reducing FPs to 37, leading to substantially higher NAB scores of 20.94 and 26.24 under the two cost profiles. The detected anomalies results further show that \cloudens captures additional anomalies, specifically Anomalies~7 and~9 in the \textit{Instant Messenger} category and Anomaly~3 in the \textit{Issue Tracker} category which are not detected by the GRU model. These findings demonstrate that \cloudens improves both FP reduction and anomaly coverage, and they highlight that different scoring strategies can reveal different types of anomalies, underscoring the challenge of evaluating anomaly detection methods in \LCSs.

\noindent\textbf{Qualitative Analysis.} Fig.~\ref{fig:5xx_count_mts} visualizes the \texttt{5xx count} subset, aggregated across all dimensions, together with the ground-truth anomalies. Fig.~\ref{fig:5xx_count_zoom_in} provides a qualitative analysis between the GRU baseline and \cloudens on three representative anomalies from the \texttt{5xx count} subset, including Anomalies 3, 9, and 17. The rows show, in order, the original telemetry, forecasting errors, and anomaly scores from both scoring strategies, along with the corresponding timestamps.

The results show that incorporating context-aware graph modeling enables \cloudens to generate forecasting errors that more clearly distinguish anomalous behavior from normal operational patterns. For Anomaly~3, only \cloudens identifies the anomaly: its forecasting errors align closely with the anomalous region (Fig.~\ref{fig:5xx_count_zoom_in}(d)), producing a clearer anomaly score under the MD scoring strategy and successful detection at the selected threshold. GRU, in contrast, produces high anomaly scores in normal regions, resulting in a missed detection under the same strategy. A more significant improvement is observed for Anomaly~9, where \cloudens successfully detects the anomaly using both scoring strategies (Figs.~\ref{fig:5xx_count_zoom_in}(h) and (k)), while GRU's forecasting errors remain insufficiently discriminative (Fig.~\ref{fig:5xx_count_zoom_in}(e)). For Anomaly~17, both models identify the anomaly under both scoring strategies, but \cloudens detects it earlier, as indicated by the green marker, contributing directly to the higher NAB scores reported in Table~\ref{tab:graph_improve}.

These visualizations demonstrate that context-aware graph modeling improves anomaly detection from two complementary perspectives. First, it produces more discriminative forecasting errors by exploiting structural relationships among cloud services. Second, the resulting anomaly scores exhibit clearer separation between normal and anomalous behavior, enabling earlier and more reliable detection of localized service failures. These findings support the design of \cloudens, which explicitly incorporates contextual information to model dependencies among distributed cloud services.

\rqanswer{EQ$_1$}{
Context-aware graph modeling significantly improves anomaly detection in \LCS monitoring, outperforming a GRU-based model in NAB score, false positive reduction, and anomaly coverage, while enabling earlier detection of localized service failures through more discriminative forecasting errors.
}

\begin{figure*}[t]
    \centering
    \includegraphics[width=\textwidth]{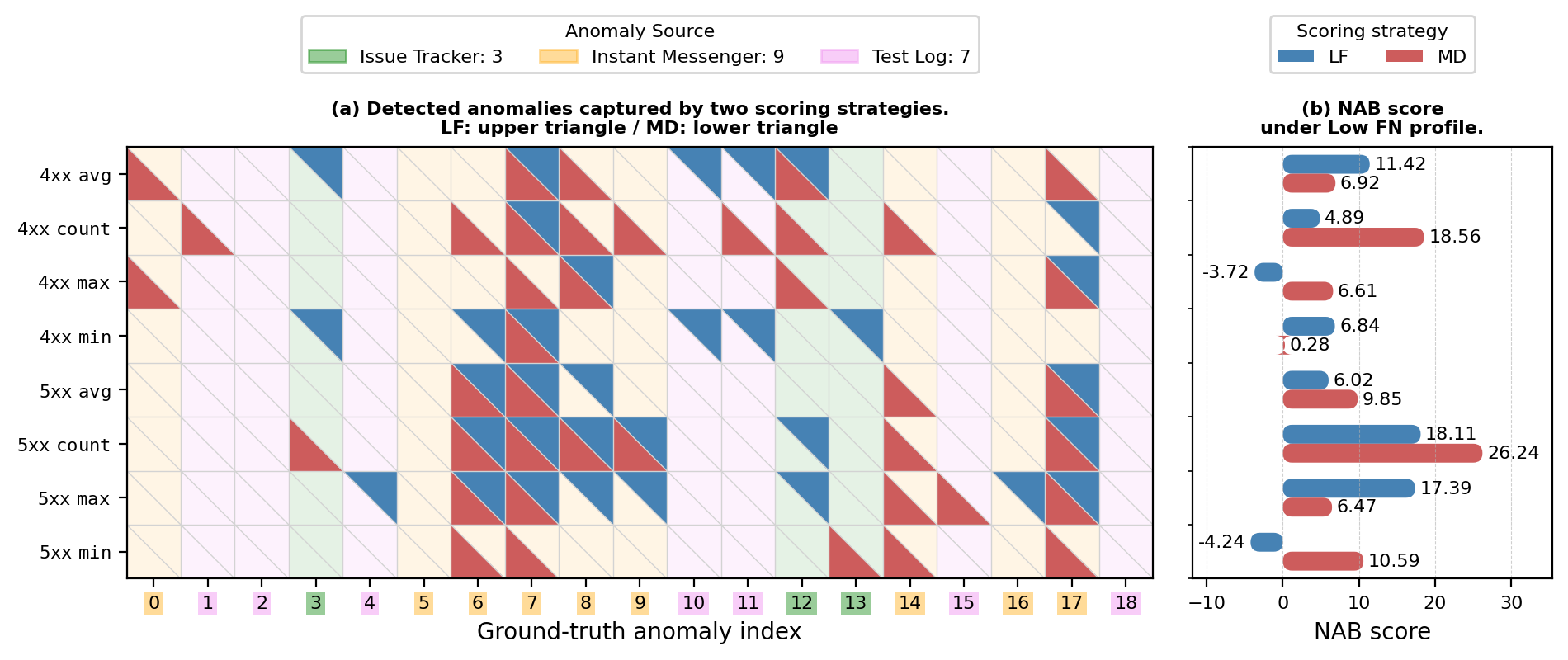}
    \caption{Detected anomalies captured by each subset.}
    \label{tab:group_other_features}
\end{figure*}

\input{tables/acm_table_3_ensemble}

\mysubsubsection{\textit{\textbf{EQ$_2$}}}{\rqsecond}
To understand how different telemetry subsets influence anomaly detection performance, we independently evaluate each subset using three telemetry sparsity imputation strategies (zero, mean, and median) together with two anomaly scoring strategies, denoted as LF and MD. Fig.~\ref{tab:group_other_features} summarizes the results obtained using the optimal configurations (see Table~\ref{tab:scoring_hyperparameters}). Specifically, Fig.~\ref{tab:group_other_features}(a) illustrates the anomalies detected by each telemetry subset using the LF and MD scoring strategies, while Fig.~\ref{tab:group_other_features}(b) reports the corresponding NAB scores under the Low FN profile. The complete experimental results for all evaluated configurations are available in our replication package~\cite{replicationPackage}.

Our experimental results reveal that anomaly detection performance varies substantially across telemetry subsets, indicating that different telemetry metrics capture complementary aspects of cloud system behavior. As shown in Fig.~\ref{tab:group_other_features}(b), the \texttt{5xx count} and \texttt{4xx count} subsets consistently achieve the highest NAB scores under the Low FN profile, reaching 26.24 and 18.56 by using MD scoring strategy, respectively. These results have been presented in Table~\ref{tab:graph_improve} and suggest that \texttt{count}-based metrics, which directly quantify the frequency of server- and client-side errors, are particularly effective for identifying obviously abnormal behaviours. 
Although the \texttt{5xx count} and \texttt{4xx count} subsets achieve the highest NAB scores, Fig.~\ref{tab:group_other_features}(a) shows that they detect only 8 and 9 of the 19 ground-truth anomalies during the testing period, respectively, indicating that the \texttt{count}-based subsets alone are insufficient to capture all anomaly patterns.
The complementary role of the remaining telemetry subsets becomes evident from Fig.~\ref{tab:group_other_features}(a), where several anomalies are detected exclusively by the \texttt{avg}, \texttt{min}, and \texttt{max} subsets. For example, with the LF scoring strategy, the \texttt{5xx max} subset uniquely detects Anomalies~4 and~16, whereas the \texttt{4xx avg} and \texttt{4xx min} subsets identify Anomalies~3, 10, and~11. 

Fig.~\ref{tab:group_other_features}(b) further shows that the effectiveness of the anomaly scoring strategy depends on the underlying telemetry subset. Specifically, the MD scoring strategy achieves higher NAB scores for several subsets, including \texttt{4xx count}, \texttt{5xx count}, and \texttt{5xx avg}, while LF performs better on the remaining subsets such as \texttt{4xx avg} and \texttt{5xx max}. Moreover, the two anomaly scoring strategies capture different anomaly patterns within the same telemetry subset. For example, on the \texttt{4xx count} subset, MD successfully identifies 8 ground-truth anomalies spanning all three anomaly sources, whereas LF detects only Anomalies~7 and~17.

These findings indicate that the performance of different telemetry subsets and anomaly scoring strategies varies considerably in terms of NAB score and anomaly coverage, providing complementary perspectives on cloud system behavior. Motivated by these observations, \cloudens exploits the complementary detection capabilities of different telemetry subsets through a subset-based ensemble strategy.
Table~\ref{tab:ensemble_configurations} summarizes the performance of different ensemble configurations, highlighting the trade-off between anomaly coverage and false alarms. The two-subset ensemble combining \texttt{4xx count} and \texttt{5xx count} achieves the highest NAB score under the Low FN profile (22.28), while detecting 10 of the 19 ground-truth anomalies, including 2 of 3 anomalies from the \textit{Issue Tracker}, 6 of 9 from the \textit{Instant Messenger}, and 2 of 7 from the \textit{Test Log}. This configuration also maintains relatively low false alarms (153 false positives) and generates only 2.01 alerts per day, representing the best balance between anomaly coverage and the increase of false alarm volume. Another two-subset combination, \texttt{4xx avg} and \texttt{5xx count}, achieves a lower NAB score but additionally detects Anomaly~0, while a three-subset combination (\texttt{4xx avg}, \texttt{5xx avg}, and \texttt{5xx max}) detects Anomaly~15.
Increasing the ensemble size improves anomaly coverage but gradually increases the alert volume. For example, combining four complementary subsets (\texttt{4xx avg}, \texttt{5xx avg}, \texttt{5xx count}, and \texttt{5xx min}) successfully detects all three \textit{Issue Tracker} anomalies and 7 of 9 \textit{Instant Messenger} anomalies. However, the number of false positives increases from 153 to 225, reducing the NAB score under the Low FN profile from 22.28 to 7.09 and increasing the average alert rate from 2.01 to 2.78 alerts per day. Replacing \texttt{4xx avg} with \texttt{4xx max} yields a similar alert volume and the same detected anomaly set, but a lower NAB score (e.g., 6.78 under the Low FN profile), underscoring the need for careful subset selection in post-processing with ensembling.

The complete subset ensemble, which combines all eight telemetry subsets (referred to as ``All'' in Table~\ref{tab:ensemble_configurations}), achieves the broadest anomaly coverage by detecting 13 of the 19 ground-truth anomalies, including all anomalies from the \textit{Issue Tracker}, 7 of 9 from the \textit{Instant Messenger}, and 3 of 7 from the \textit{Test Log}. This improvement comes with a significant trade-off, increasing the number of false positives to 402 and the average alert volume to 4.98 alerts per day, while reducing the NAB score under the Low FN profile to -10.05.

\rqanswer{EQ$_2$}{
Detection performance varies substantially across telemetry subsets and anomaly scoring strategies. The \texttt{4xx count} and \texttt{5xx count} subsets achieve the highest NAB scores, detecting 9 and 8 of the 19 ground-truth anomalies, respectively. Other subsets reveal complementary anomaly patterns despite lower NAB scores. Additionally, the ensemble approach provides a balance between anomaly coverage and false alerts.
}

\begin{figure}[!ht]
    \centering
    \includegraphics[width=\columnwidth]{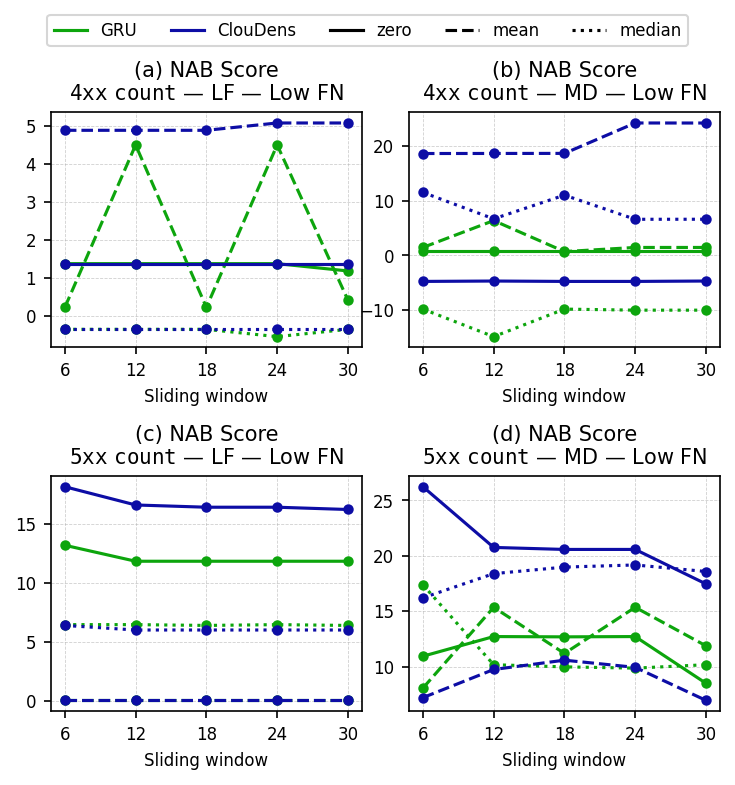}
    \caption{Impact of sliding windows and telemetry sparsity imputation on NAB Score.}
    \label{fig:nab_score_across_sliding_windows}
    \vspace{-10pt}
\end{figure}

\begin{figure}[!ht]
    \centering
    \includegraphics[width=\columnwidth]{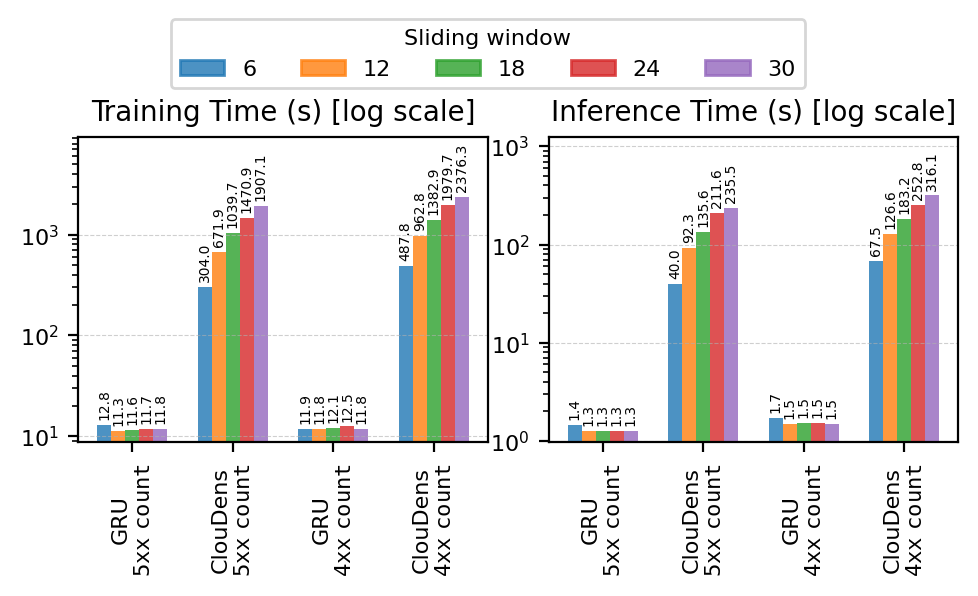}
    \caption{Impact of sliding windows on Training/Inference Time.}
    \label{fig:training_inference_time}
     \vspace{-10pt}
\end{figure}
\mysubsubsection{\textit{\textbf{EQ$_3$}}}{\rqthird}\hfill

\noindent\textbf{Detection Performance.}
Fig.~\ref{fig:nab_score_across_sliding_windows} illustrates the impact of sliding window size and sparsity imputation strategy on the NAB score of \cloudens and GRU for the \texttt{4xx count} and \texttt{5xx count} subsets under the Low FN profile. Several observations emerge.
Overall, detection performance varies considerably with the imputation strategy, and the best strategy differs by subset: on \texttt{4xx count}, \cloudens performs best with mean imputation (Fig.~\ref{fig:nab_score_across_sliding_windows}(a) and (b)), whereas on \texttt{5xx count}, it performs best with zero imputation across nearly all window sizes (Fig.~\ref{fig:nab_score_across_sliding_windows}(c) and (d)).

For the \texttt{5xx count} subset, \cloudens substantially outperforms GRU under zero imputation. Under the LF scoring strategy, \cloudens achieves an NAB score of 18.11 versus 13.16 for GRU at a window size of 6 (consistent with Table~\ref{tab:graph_improve}), and this gap remains stable as window size increases. For the other imputation strategies, the two models show only small differences. Both models maintain stable NAB scores across window sizes under the LF strategy; under the MD strategy, \cloudens continues to outperform GRU with zero imputation, though its score drops sharply at a window size of 30. GRU shows modest gains with mean imputation but fluctuates considerably across window sizes, indicating unstable performance.
For the \texttt{4xx count} subset, \cloudens remains stable under both scoring strategies, while GRU fluctuates substantially under mean imputation with the LF strategy; both models are stable under the MD strategy. Mean imputation yields the best results for this subset, where \cloudens clearly outperforms GRU, reaching its highest NAB score of 24.11 at a window size of 30.

Taken together, these results show that the optimal imputation strategy for telemetry sparsity is subset-dependent rather than universal, and that \cloudens consistently delivers more stable NAB scores than GRU once paired with its best-performing imputation strategy for each subset. This highlights the need for careful empirical tuning when evaluating detection performance in \LCSs, as sparsity imputation, sliding window sizes, and scoring functions can substantially affect results.

\noindent\textbf{Computation Costs.}
Fig.~\ref{fig:training_inference_time} compares training and inference time between the GRU forecasting model and \cloudens across different sliding window sizes, using the \texttt{5xx count} and \texttt{4xx count} as representative subsets. The remaining subsets (\texttt{avg}, \texttt{min}, and \texttt{max}) share the same graph structure and dimensionality as their corresponding \texttt{count} subset within each status code category, and therefore exhibit nearly identical runtime behavior; we omit them here for brevity.

As expected, GRU consistently trains and infers faster than ClouDens, since the latter's attention mechanisms and graph-based operators add computational overhead. GRU-based models require about 11.3–12.8 seconds for training and 1.3–1.4 seconds for inference on the \texttt{5xx count} subset, and similar training and inference times on \texttt{4xx count}, regardless of window size. \cloudens, however, scales roughly linearly with window size (6 to 30 timestamps): training time rises from 304.0 to 1907.1 seconds on \texttt{5xx count} and 487.8 to 2376.3 seconds on \texttt{4xx count}, while inference time rises from 40.0 to 235.5 seconds and 67.5 to 316.1 seconds, respectively.
Given a 26,488-timestamp testing period, per-timestamp inference time ranges from about 1.5 ms in the best case to 12 ms in the worst case. Thus, despite its higher computational cost, per-timestamp inference overhead of \cloudens remains practical for cloud service monitoring, where improved detection quality justifies the added cost.

\rqanswer{EQ$_3$}{Across varying sliding window sizes, \cloudens consistently achieves higher and more stable NAB scores than GRU once paired with its best-performing sparsity imputation strategy for each subset, though this comes at a proportionally higher computational cost for both training and testing stages.
}

\section{Discussion\label{sec:threats}}
Despite the promising results achieved by \cloudens, several practical aspects deserve further consideration to guide its application in real-world cloud environments.

\textit{\textbf{Practical Insights.}}
The experimental results provide several practical insights for designing anomaly detection frameworks in \LCSs. First, incorporating operational-context  attributes encoded in API activity telemetry substantially improves detection performance by enabling forecasting models to exploit structural dependencies among distributed cloud services (as shown in \textbf{EQ$_1$}). This results in more discriminative anomaly scores, higher detection accuracy, and earlier detection than purely temporal models.

Second, different telemetry subsets capture complementary aspects of operational behaviors in \LCSs. \texttt{count}-based metrics are generally more effective for identifying clear anomaly patterns, whereas other statistical aggregations (\texttt{avg}, \texttt{min}, and \texttt{max}) reveal anomalies that may be missed by \texttt{count}-based subsets alone. Moreover, detection performance is highly sensitive to telemetry sparsity imputation and anomaly scoring strategies (as shown in \textbf{EQ$_2$} and \textbf{EQ$_3$}). This sensitivity poses a major challenge in anomaly detection for \LCSs, making fair comparisons difficult. Consequently, benchmarks for \LCSs should systematically evaluate these design choices rather than relying on a single evaluation setting.

Finally, combining complementary telemetry subsets through an ensemble strategy improves anomaly coverage, particularly under evaluation metrics such as the NAB score, which rewards early detection and supports different operational priorities through configurable cost profiles. However, the ensemble post-processing also raises the number of false positives. Therefore, effective ensemble design should prioritize complementary telemetry subsets that maximize anomaly coverage while maintaining an acceptable alert volume, rather than simply aggregating all available detectors (as shown in \textbf{EQ$_2$}).

\textit{\textbf{Limitations.}} Although \cloudens demonstrates significant improvements for \texttt{count}-based telemetry subsets, several limitations still remain. First, the current framework relies on manually designed telemetry feature decomposition and context-aware graph construction, which may require domain knowledge when applied to other cloud platforms. 
Second, selecting appropriate imputation strategies for telemetry sparsity, anomaly scoring functions, and ensemble configurations still requires empirical investigation. Third, we employ A3T-GCN only to demonstrate the benefit of operational-context attributes for monitoring services in \LCSs, not to establish it as the optimal backbone; other graph-based models have not been investigated. Since detection performance varies considerably across telemetry subsets, imputation strategies, and scoring strategies, a systematic benchmark of forecasting backbones in the context of monitoring services in \LCSs remains important to discover. We leave these limitations for future research.

\textit{\textbf{Threats to Validity.}} Despite careful experimental design, this study is subject to several threats to validity~\cite{wohlin2012experimentation,yin2009case}.

\begin{itemize}

\item \textit{Internal Validity:} Our results may be affected by model configuration, hyperparameter tuning, and threshold selection. Although we evaluated multiple configurations for \cloudens, the search space was not exhaustive, and alternative settings may yield different results. To facilitate verification and replication, we publicly release our implementation and experimental results~\cite{replicationPackage}.

\item \textit{External Validity:} The evaluation is conducted on a single real-world IBM Cloud Telemetry Dataset. Although the dataset represents a production cloud environment and has been recognized as a critical case study~\cite{islam2025anomaly,yin2009case}, our findings may not generalize to all cloud platforms or telemetry sources. Nevertheless, the proposed methodology and evaluation protocol are applicable to other cloud telemetry datasets.

\item \textit{Construct Validity:} We evaluate anomaly detection using the NAB score and confusion-matrix-based metrics. While these metrics are widely adopted, the completeness of ground-truth anomaly labels may affect the measured performance, particularly for subtle or previously unknown anomalies.

\end{itemize}

\section{Conclusions\label{sec:conclusion}}
Building reliable and effective anomaly detection solutions in the context of cloud network monitoring remains challenging. This is clearly reflected in the recent IBM Cloud Telemetry Dataset released by Islam \etal~\cite{islam2025anomaly}, which exhibits three key characteristics: high-dimensional telemetry performance metrics, complex temporal and spatial dependencies among distributed cloud services, and severe telemetry sparsity.
Motivated by these observations, we conducted a deeper empirical investigation of the dataset to better understand the challenges of cloud network monitoring with telemetry logs and developed \cloudens, an operational context-aware anomaly detection framework for \LCSs. To deal with these challenges, \cloudens integrates telemetry feature decomposition, operational-context graph modeling, and spatio-temporal forecasting, and further examines different post-processing techniques, including two scoring strategies and ensembling. Our experimental results demonstrate that \cloudens outperforms a GRU-based forecasting baseline by achieving higher NAB scores, broader anomaly coverage, and earlier detection.

\textbf{Practical Guidelines}: 
Our empirical study shows that telemetry feature subsets, operational-context modeling, telemetry sparsity imputation, anomaly scoring strategies, and ensemble configurations all substantially influence anomaly detection performance in \LCSs. These findings demonstrate that the effectiveness of anomaly detection depends not only on the forecasting model but also on telemetry representation and post-processing design choices, which should therefore be systematically considered when designing and benchmarking anomaly detection methods for \LCS monitoring.

\textbf{Future Work.} The empirical study highlights several opportunities for future research. Promising directions include extending \cloudens to investigate dynamic graph structures that adapt to evolving service topologies, developing adaptive methods for selecting anomaly scoring strategies and ensemble configurations, and integrating root-cause localization to facilitate automated fault diagnosis. Furthermore, evaluating \cloudens across additional cloud platforms and real-time deployment scenarios would provide a broader assessment of its practical applicability for \LCS monitoring.

\textbf{Reproducible package.} Our code and experimental results are publicly available at~\cite{replicationPackage}.

\newpage
\section*{Acknowledgements}
This work has been partially funded by 
(a) the MUR (Italy) Department of Excellence 2023 - 2027, 
(b) the European HORIZON-KDT-JU research project MATISSE ``Model-based engineering of Digital Twins for early verification and validation of Industrial Systems", HORIZON-KDT-JU-2023-2-RIA, Proposal number:  101140216-2, KDT232RIA\_00017,
(c) the SFI Smart Ocean project (\url{https://sfismartocean.no}).

\ifCLASSOPTIONcaptionsoff
  \newpage
\fi

\bibliographystyle{IEEEtran}
\bibliography{references}

\end{document}

%% file: tables/acm_table_1_graph_improvement.tex
 \begin{table*}
   \caption{Comparison of \cloudens and GRU on the \textbf{``5xx count''} feature subset with a sliding window size of 6, using the Likelihood Function and Mahalanobis Distance scoring strategies.
   }
   \label{tab:graph_improve}
\resizebox{\textwidth}{!}{%
    \begin{threeparttable}
  \begin{tabular}{rrccccccccc} 
    \toprule
    
    \multirowcell{3}[0pt][r]{Scoring\\Strategy}&\multirowcell{3}[0pt][r]{Model}&\multicolumn{4}{c}{Confusion Matrix\tnote{a}}&\multicolumn{2}{c}{NAB Score}&\multicolumn{3}{c}{Detected Anomalies\tnote{b}} \\
    \cmidrule(lr){3-6}
    \cmidrule(lr){7-8}
    \cmidrule(lr){9-11}
    
     & & \multirow{2}{*}{TP} & \multirow{2}{*}{TN} & \multirow{2}{*}{FP} & \multirow{2}{*}{FN} &\multirowcell{2}{Standard\\Profile}& \multirowcell{2}{Low FN\\Profile}& \multirowcell{2}{Issue Tracker} & \multirowcell{2}{Instant Messenger} & \multirowcell{2}{Test Log}\\
     & &  &  &  &  &  & & & &\\
    \toprule

    \multirowcell{2}[0pt][r]{Likelihood Function\\$\{W,W',L_t\}=\{30,2,0.99975\}$}&\multirowcell{2}[0pt][r]{GRU }& 6 & 25468 & 53 & 961 & 6.58 & 13.16 & 1/3 & 4/9 & 0/7 \\
    & &  &  &  &  &  &  & [12] & [6, 7, 8, 17] &\\
    \cmidrule{2-11}
    
    \multirowcell{2}[0pt][r]{Mahalanobis Distance\\$\{\epsilon\}=\{99.8\}$}&\multirowcell{2}[0pt][r]{GRU }& 13 & 25481 & 40 & 954 & 5.89 & 10.95 & 0/3 & 4/9 & 0/7 \\
    &  &  &  &  &  &  &  & 0 & [6, 8, 14, 17] &\\
    \midrule
    
    \multirowcell{2}[0pt][r]{Likelihood Function\\$\{W,W',L_t\}=\{30,2,0.99975\}$}&\multirowcell{2}[0pt][r]{\cloudens}& 7 & 25469 & 52 & 960 & \textbf{11.38} & \textbf{18.11} & 1/3 & 5/9 & 0/7\\
    &  &  &  &  &  &  &  & [12] & [6, 7, 8, \textbf{9}, 17] &\\
    \cmidrule{2-11}

    \multirowcell{2}[0pt][r]{Mahalanobis Distance\\$\{\epsilon\}=\{99.8\}$}&\multirowcell{2}[0pt][r]{\cloudens}& 16 & 25483 & 37 & 952 & \textbf{20.94} & \textbf{26.24} & 1/3 & 6/9 & 0/7\\
    &  &  &  &  &  &  &  & [\textbf{3}] & [6, \textbf{7}, 8, \textbf{9}, 14, 17] &\\
    \hline

    \bottomrule
\end{tabular}
\begin{tablenotes}
    \item[a] The confusion matrix reports point-based results. \tnote{b} A ground-truth anomaly is considered detected if at least one TP is identified within its window.
\end{tablenotes}
\vspace{-10pt}
\end{threeparttable}
}
 \end{table*}

%% file: tables/acm_table_3_ensemble.tex
\begin{table*}[t]
\caption{Performance of \cloudens with different ensemble configurations using \colorbox{IndianRed!50}{Mahalanobis Distance}.}
\label{tab:ensemble_configurations}
\resizebox{\textwidth}{!}{%

\begin{threeparttable}
\begin{tabular}{llccccllcccc}
\toprule
\multirowcell{3}[0pt][l]{\#\\Sub.} & \multirowcell{3}[0pt][l]{List of subsets} &\multicolumn{4}{c}{Confusion Matrix}& \multicolumn{2}{c}{NAB Score} & \multicolumn{3}{c}{Detected Anomalies} & \multirowcell{3}[0pt][l]{Avg. Alerts\\per day\\ \textit{(92 days)}} \\
\cmidrule(lr){3-6}
\cmidrule(lr){7-8}
\cmidrule(lr){9-11}
 &  &  \multirowcell{2}[0pt][l]{TP} & \multirowcell{2}[0pt][l]{TN} & \multirowcell{2}[0pt][l]{FP} & \multirowcell{2}[0pt][l]{FN} &  \multirowcell{2}[0pt][c]{Standard\\Profile} & \multirowcell{2}[0pt][c]{Low FN\\Profile}  &  \multirowcell{2}[0pt][c]{Issue\\Tracker} & \multirowcell{2}[0pt][c]{Instant Messenger} & \multirowcell{2}[0pt][c]{Test Log} &  \\
 &  &  &  &  &  &  &  &  &  &  &   \\
\toprule
2 & \multirowcell{2}[0pt][l]{\texttt{4xx avg}, \texttt{5xx count}} & 24 & 25387 & 134 & 943 & 5.46 & 19.43 & \multirowcell{2}[0pt][c]{2/3 \\ \text{[3, 12]}} & \multirowcell{2}[0pt][c]{7/9 \\ \text{[\textbf{0}, 6, 7, 8, 9, 14, 17]}} & \multirowcell{2}[0pt][c]{0/7 \\ \text{[]}} & 1.72 \\
 &  &  &  &  &  &  &  &  &  &  &  \\
\midrule
2 & \multirowcell{2}[0pt][l]{\texttt{4xx count}, \texttt{5xx count}} & 32 & 25368 & 153 & 935 & \textbf{7.11} & \textbf{22.28} & \multirowcell{2}[0pt][c]{2/3 \\ \text{[3, 12]}} & \multirowcell{2}[0pt][c]{6/9 \\ \text{[6, 7, 8, 9, 14, 17]}} & \multirowcell{2}[0pt][c]{2/7 \\ \text{[1, 11]}} & 2.01 \\
 &  &  &  &  &  &  &  &  &  &  &  \\
\midrule
3 & \multirowcell{2}[0pt][l]{\texttt{4xx avg}, \texttt{5xx avg},\\ \texttt{5xx max}} & 24 & 25330 & 191 & 943 & -15.34 & 3.81 & \multirowcell{2}[0pt][c]{1/3 \\ \text{[12]}} & \multirowcell{2}[0pt][c]{6/9 \\ \text{[0, 6, 7, 8, 14, 17]}} & \multirowcell{2}[0pt][c]{1/7 \\ \text{[\textbf{15}]}} & 2.34 \\
 &  &  &  &  &  &  &  &  &  &  &  \\
\midrule
4 & \multirowcell{2}[0pt][l]{\texttt{4xx avg}, \texttt{5xx avg},\\ \texttt{5xx count}, \texttt{5xx min}} & 31 & 25296 & 225 & 936 & -15.69 & 7.09 & \multirowcell{2}[0pt][c]{3/3 \\ \text{[3, 12, 13]}} & \multirowcell{2}[0pt][c]{7/9 \\ \text{[0, 6, 7, 8, 9, 14, 17]}} & \multirowcell{2}[0pt][c]{0/7 \\ \text{[]}} & 2.78 \\
 &  &  &  &  &  &  &  &  &  &  &  \\
\midrule
4 & \multirowcell{2}[0pt][l]{\texttt{4xx max}, \texttt{5xx avg},\\ \texttt{5xx count}, \texttt{5xx min}} & 33 & 25298 & 223 & 934 & -16.15 & 6.78 & \multirowcell{2}[0pt][c]{3/3 \\ \text{[3, 12, 13]}} & \multirowcell{2}[0pt][c]{7/9 \\ \text{[0, 6, 7, 8, 9, 14, 17]}} & \multirowcell{2}[0pt][c]{0/7 \\ \text{[]}} & 2.78 \\
 &  &  &  &  &  &  &  &  &  &  &  \\
\midrule
8 & All & 56 & 25119 & 402 & 911 & -49.29 & -10.05 & \multirowcell{2}[0pt][c]{3/3 \\ \text{[3, 12, 13]}} & \multirowcell{2}[0pt][c]{7/9 \\ \text{[0, 6, 7, 8, 9, 14, 17]}} & \multirowcell{2}[0pt][c]{3/7 \\ \text{[1, 11, 15]}} & 4.98 \\
 &  &  &  &  &  &  &  &  &  &  &  \\
\hline
\bottomrule
\end{tabular}
\vspace{-15pt}
\end{threeparttable}
}
\end{table*}